\documentclass[ 
 reprint,
 showkeys,
 amsmath,amssymb,
 aps,
]{revtex4-2}

\usepackage{graphicx}
\usepackage{dcolumn}
\usepackage{svg}
\usepackage{natbib}
\usepackage{subcaption}
\usepackage{caption}
\usepackage{float}
\usepackage{enumitem} 
\makeatletter
\let\newfloat\newfloat@ltx
\makeatother
\usepackage[noend]{algpseudocode}
\usepackage{algorithm}
\usepackage{placeins}
\usepackage{bm}
\usepackage{hyperref}
\hypersetup{
    colorlinks=true,
        linktoc=all,
    linkcolor=blue,
    citecolor=blue,
    urlcolor=blue
}

\begin{document}
\preprint{APS/123-QED}

\title{Detection and Prediction of Future Massive Black Hole Mergers with Machine Learning and Truncated Waveforms}
\author{Niklas Houba}
\email{niklas.houba@erdw.ethz.ch}
\author{Stefan H. Strub}%
\author{Luigi Ferraioli}%
\author{Domenico Giardini}%
\affiliation{%
 Institute of Geophysics \\  Department of Earth and Planetary Sciences at ETH Zurich \\
 Sonneggstrasse 5, 8092 Zurich, Switzerland
}

\begin{abstract}
We present a novel machine learning framework tailored to detect massive black hole binaries observed by spaceborne gravitational wave detectors like the Laser Interferometer Space Antenna (LISA) and predict their future merger times. The detection is performed via convolutional neural networks that analyze time-evolving Time-Delay Interferometry (TDI) spectrograms and utilize variations in signal power to trigger alerts. The prediction of future merger times is accomplished with reinforcement learning. Here, the proposed algorithm dynamically refines time-to-merger predictions by assimilating new data as it becomes available. Deep Q-learning serves as the core technique of the approach, utilizing a neural network to estimate Q-values throughout the observational state space. To enhance robust learning in a noisy environment, we integrate an actor-critic mechanism that segregates action proposals from their evaluation, thus harnessing the advantages of policy-based and value-based learning paradigms. 
We leverage merger estimation obtained via template matching with truncated waveforms to generate rewards for the reinforcement learning agent. These estimations come with inherent uncertainties, which magnify as the merger event stretches further into the future. The reinforcement learning setup is shown to adapt to these uncertainties by employing a dedicated policy fine-tuning approach, ensuring the reliability of predictions despite the varying degrees of template-matching precision. The algorithm denotes a first step toward a low-latency model that provides early warnings for impending transient phenomena. By delivering timely and accurate forecasts of merger events, the framework supports the coordination of gravitational wave observations with accompanied electromagnetic counterparts, thus enhancing the prospects of multi-messenger astronomy. We use the LISA Spritz data challenge for validation.
\end{abstract}

\maketitle


\section{\label{sec:level1}Introduction}
The Laser Interferometer Space Antenna (LISA) is an advanced space mission slated for launch in the 2030s. The mission is designed to detect gravitational waves in the millihertz frequency band. In contrast to its terrestrial counterparts, like the Laser Interferometer Gravitational-Wave Observatory (LIGO) and Virgo, which detect gravitational waves ranging from 10 Hz to 10 kHz, LISA seeks to enhance these observations by probing the low-frequency spectrum spanning from 0.1 mHz to 1 Hz \cite{Martynov_2016, colpi2024lisa}. The mission's capability is expected to expand the horizons of gravitational wave astronomy, allowing for the exploration of cosmic phenomena beyond the reach of ground-based detectors \cite{SciReqDoc}.

\begin{figure}[]
	\centering
  \includegraphics[trim=7 10 380 0, clip,width=0.47\textwidth]{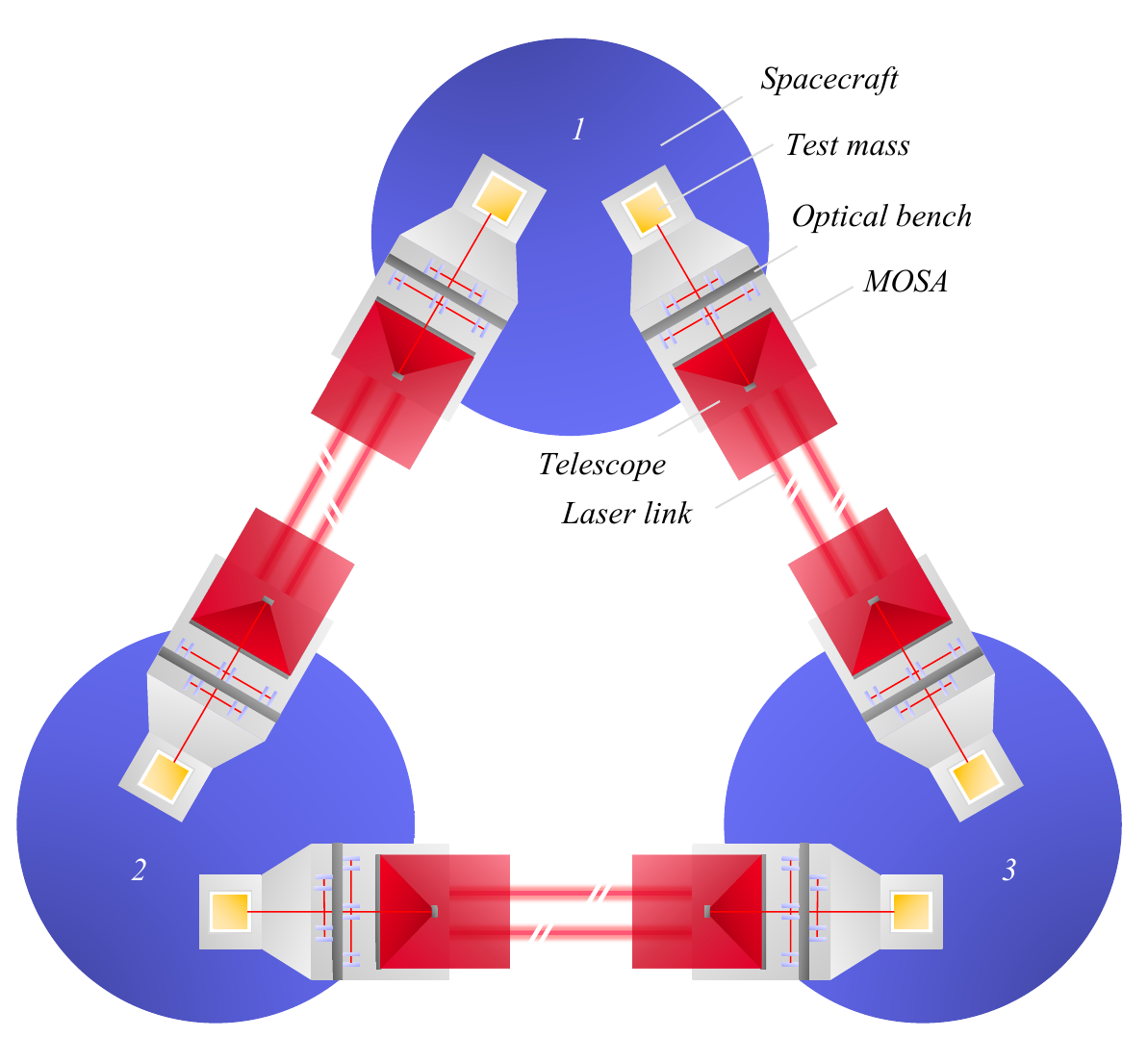}
	\caption{Configuration of the LISA constellation. The constellation comprises three spacecraft optically connected via bi-directional near-infrared laser links spanning a distance of 2.5 million kilometers. MOSA stands for moving optical subassembly. A MOSA mainly accommodates an optical bench, a telescope, and a test mass. The optical design is highly simplified.}
	\label{im:LISAConstellation}
\end{figure}

Utilizing a trio of spacecraft, LISA's concept involves monitoring the relative variations in distance between free-falling test masses induced by the passage of gravitational waves. Positioned in an approximately equilateral triangle configuration extending over 2.5 million kilometers, the three spacecraft orbit the Sun in an Earth-trailing or Earth-leading formation \cite{Martens2021}. Laser interferometry is employed to precisely measure the relative distance variations between the free-falling test masses housed within each spacecraft. These test masses need to be rigorously shielded from external forces and disturbances. Therefore, the spacecraft serve as protective enclosures, maneuvering around the test masses and adjusting their positions to maintain the free-fall condition along the sensitive measurement axes \cite{BenderArt}. A schematic illustration of the configuration is given in Fig. \ref{im:LISAConstellation}.

Gravitational waves passing through the LISA constellation slightly perturb the distances between the test masses through spacetime stretching and compression \cite{Shaddock2008}. These alterations are exceedingly subtle, on the scale of fractions of an atom's diameter across the vast distances separating the spacecraft \cite{MPGWebsite1}. Nevertheless, by analyzing the interference patterns generated by the laser beams as they traverse between the spacecraft, LISA will be capable of detecting these minuscule changes in distance on a picometer level \cite{Schuldt_2009}.

One of the primary astrophysical sources of interest for LISA are merging massive binary black holes from $10^5$ to $10^7\,\mathrm{M}_{\odot}$ up to redshift $z\sim10$ \cite{Mangiagli2022}. These events produce gravitational waves that carry information about the dynamics and environments of their sources and also serve as probes for testing theories of gravity in strong gravitational fields. The detections are expected to provide insights into the formation of galaxies. By observing the mergers from massive black hole systems, LISA could also give perspectives about the nature of dark matter and dark energy \cite{Barausse2015}.

\begin{figure}[b]
	\centering
  \includegraphics[trim=60 60 0 10, clip,width=0.47\textwidth]{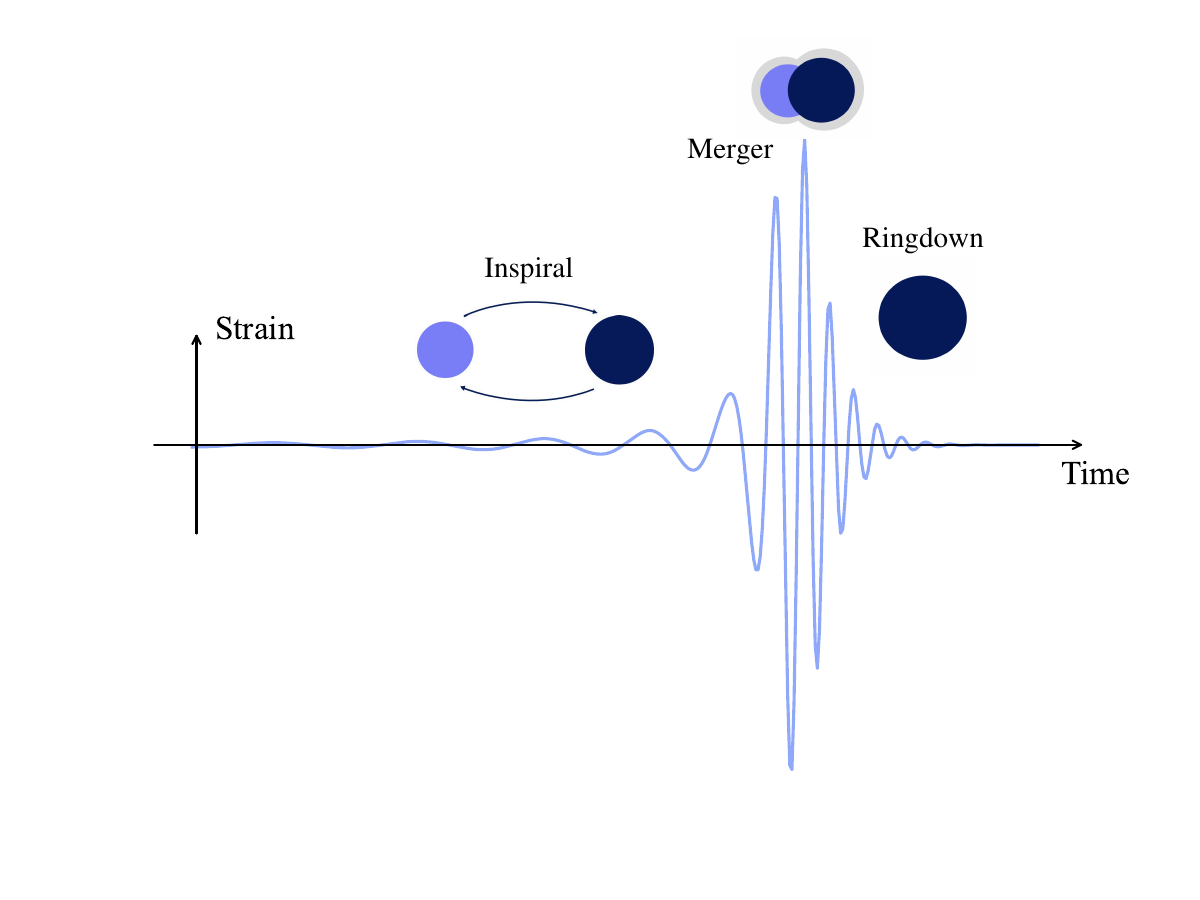}
	\caption{Qualitative illustration of the progression through the inspiral, merger, and ringdown phases of a massive black hole binary system. The initial frequency of the gravitational wave is approximately double the Keplerian frequency. With energy loss due to gravitational wave emission, the frequency increases over time, resulting in a distinctive chirping signal. The paper seeks to detect the signal and anticipate the merger time prior to its occurrence by analyzing temporal and spectral variations in signal power during the inspiral.}
	\label{im:MBHBQualitative}
\end{figure}

\noindent
The gravitational strains emitted by massive binary black hole mergers are complex. These events typically unfold through three phases: the inspiral, where the black holes draw closer while orbiting each other; the merger, during which the black holes merge into a single entity; and the ringdown, a period in which the newly formed black hole stabilizes into its final state. Each phase has distinctive waveform characteristics influenced by the masses and spins of the black holes involved. A qualitative illustration of the produced gravitational strain is given in Fig. \ref{im:MBHBQualitative}. Understanding the gravitational waveform during the inspiral phase is crucial for detecting these phenomena as early as possible and predicting their evolution precisely. The algorithm proposed in this paper aims to identify temporal and spectral changes in signal power as part of the detection process.

\subsection{\label{ssec1:level1} Early Warning and Alert Systems}
Early warning systems play a vital role across various domains where timely predictions of significant events facilitate preemptive measures or preparatory actions. In seismology, for example, algorithms analyze real-time seismic data to forecast earthquakes \cite{Bose2008}, providing crucial preparation time to enhance response efforts and potentially mitigate casualties. The effectiveness of these systems relies on their ability to accurately interpret precursor signals and issue reliable alerts before the event unfolds completely. The underlying methodology, time-to-event prediction or survival analysis \cite{ling2023learning, Shahin2024, Wang2022}, is utilized in other contexts as well, such as medical research for predicting patient survival rates and financial markets for assessing credit risk \cite{Huang2023, CreditRiskAssessment}. What remains consistent across the diverse applications is the accommodation and analysis of right-censored data; data where the event of interest has not yet transpired.

While LIGO has successfully implemented early warning systems, generating alerts shortly before mergers of spiraling neutron stars \cite{LIGOWebsite1}, analogous low-latency data analysis and alert pipelines for LISA remain a subject of ongoing research and development. Drawing inspiration from LIGO's advancements, 
the paper strengthens the existing data processing and analysis frameworks by introducing a dedicated detection and prediction pipeline tailored for impending massive black hole mergers.
With LISA's sensitivity to low-frequency gravitational waves, the observatory is uniquely positioned to record the extended inspiral phase of these events. This capability spans from hours to days or even weeks before the culmination of the merger, varying with the source's specific parameters, such as component masses and luminosity distance \cite{PhysRevD.102.084056}. This extended observation window is central for the development of the detection and prediction pipeline, which aims to provide the astronomical community with significant lead times. Such advance alerts are crucial for organizing multi-messenger campaigns \cite{PhysRevD.108.103034}, enhancing the scientific value derived from each observation.

\subsection{\label{ssec3:level2} Challenges in LISA Data Analysis}
Developing predictive algorithms for LISA presents unique challenges stemming from the mission's target nature and the technical complexities involved in space-based gravitational wave detection and data analysis. LISA's measurements are subject to various noise sources, including instrumental disturbances intrinsic to the spacecraft constellation \cite{PhysRevD.105.122008, hartig2021nongeometric, hartig2022geometric}, transient noise artifacts \cite{Robson2019, PhysRevD.106.042006, houba2024detection}, and environmental perturbations \cite{Armano2015, Frank2020}. The interferometric measurements obtained from LISA necessitate complicated pre-processing before astrophysical data analysis becomes possible \cite{Houba2023, https://doi.org/10.15488/8545}. 

The complexity of data analysis is further compounded by the expected volume of detectable sources. With thousands of gravitational wave sources anticipated to overlap temporally and spectrally \cite{PhysRevD.104.043019}, LISA's global data analysis framework must disentangle these concurrent signals to isolate individual events for study. While this objective is different from the task of an early warning pipeline, the overlapping signals scenario presents a significant challenge that the system must deal with. 

Overall, developing predictive models for LISA requires a multidisciplinary approach bridging astrophysics, data science, computational mathematics, and software engineering. Collaboration across these disciplines is essential for innovating solutions to the technical and theoretical challenges inherent in gravitational wave prediction. The paper marks a first step toward fostering such collaboration.

\subsection{\label{ssec3:level2}Overview of the Paper}
The primary aim of this paper is to develop a prototype pipeline for the near real-time detection of massive black hole binaries and the prediction of their future merger times for early warning and alert purposes. 
The study focuses on identifying changes in the detected signal power and recognizing temporal patterns within the measured signals considering quasi-continuous data updates.
Section \ref{sec:MBHBWaveforms} presents the gravitational waveform of massive black hole binaries, detailing the challenges involved in detecting these signals during the inspiral phase amid the realistic measurement noise expected for LISA. 
Section \ref{sec:levelDet} describes the detection methodology utilizing a sequence of convolutional neural networks with spectrograms as input. The section elaborates on the signal detection results and investigates the capabilities of supervised learning for massive binary black hole detection across varied observational conditions.
Section \ref{sec:level2} discusses the formalism of Markov decision processes (MDPs) in the context of reinforcement learning and summarizes the mathematical principles of temporal difference learning and Q-learning. Section \ref{sec:RLFL} derives the reinforcement learning framework for time-to-merger prediction in LISA denoting the second step of the proposed pipeline.
Section \ref{sec:level3} explores the utilization of truncated waveform models to generate rewards for the reinforcement learning agent via template matching. The section examines the impact of contaminated rewards on prediction accuracy and proposes a supervised learning-based policy fine-tuning approach to counteract this issue. The prediction results obtained from this approach are presented.
Section \ref{sec:level5} elaborates on the testing and validation procedures, utilizing the LISA Spritz dataset as an independent benchmark to assess the predictive capabilities of the developed pipeline.
Section \ref{sec:level6} summarizes the results and explores potential avenues for future investigations.

\section{Gravitational Waveforms of Massive Black Hole Binaries}\label{sec:MBHBWaveforms}
Gravitational waves emitted by massive black hole binaries provide information about the dynamics of the inspiral, merger, and ringdown phases through their characteristic waveforms. Models of these waveforms try to describe the temporal evolution of the amplitude and phase of the gravitational wave that can be utilized to detect and analyze those events in the complex LISA data. Accurate waveform models are essential for interpreting gravitational wave signals, necessitating advanced computational techniques that encompass both analytical and numerical methods to solve Einstein's field equations under special assumptions and across different stages of binary evolution.

Model generation for the inspiral phase of binary black hole systems often relies on post-Newtonian theory, an analytical approximation of general relativity that expands the equations of motion in powers of $(v/c)$, where $v$ is the velocity of the orbiting bodies, and $c$ is the speed of light. For the complex dynamics of the merger and ringdown phases, numerical relativity simulations provide the most accurate descriptions, modeling the full non-linearities of general relativity. These approaches ensure the generation of a reliable model, like the one termed as IMRPhenomD \cite{PhysRevD.93.044007}, which will be applied in this paper.
\subsection{The IMRPhenomD Waveform Model}
The IMRPhenomD waveform model is part of a series of phenomenological models that provide an analytical description of gravitational waveforms from binary black hole coalescences. This model is designed to accurately depict the gravitational waves generated throughout the coalescence process, which includes the inspiral, merger, and ringdown phases. IMRPhenomD is characterized by its approach to modeling waveforms for binary systems where the spin of each black hole is either aligned or anti-aligned with the orbital angular momentum to simplify the spin dynamics. The model considers the dominant (2,2) mode and is parameterized by the masses of the black holes $m_1$ and $m_2$, the dimensionless spin parameters $\chi_1$ and $\chi_2$, the luminosity distance $D$, the merger time $t_{\text{m}}$, the reference frequency $f_{\text{ref}}$ at which $t_{\text{m}}$ is set, and the reference phase $\phi_{\text{ref}}$ corresponding to the reference frequency $f_{\text{ref}}$ \cite{PhysRevD.93.044007}. IMRPhenomD incorporates elements from numerical relativity simulations into its calibration process, specifically focusing on binaries with non-precessing spins. The phase evolution during the inspiral is complemented by an effective-one-body (EOB) approach for the merger and ringdown phases, aiming to balance accuracy and computational efficiency. 
The core idea behind the EOB approach is to map the dynamics of a two-body system onto an effective problem of a single body moving in a certain effective metric. This metric encapsulates the complex gravitational interactions between the two bodies. The approach combines elements from post-Newtonian theory, black hole perturbation theory, and numerical relativity to provide a comprehensive and accurate description of the binary dynamics \cite{TiecAlexandre2014}.
The general expression for the IMRPhenomD waveform in the frequency domain is: 
\begin{equation}
h(f) = A(f)\cdot e^{i\Phi(f)},    
\end{equation}
where $A(f)$ and $\Phi(f)$ represent the amplitude and phase. The phase, in particular, draws from post-Newtonian theory and numerical relativity simulations, offering a synthesized description of the dynamics of binary black hole evolution.

\begin{figure}[]
    \centering
        \centering
        \includegraphics[trim=10 0 0 37, clip, width=0.52\textwidth]{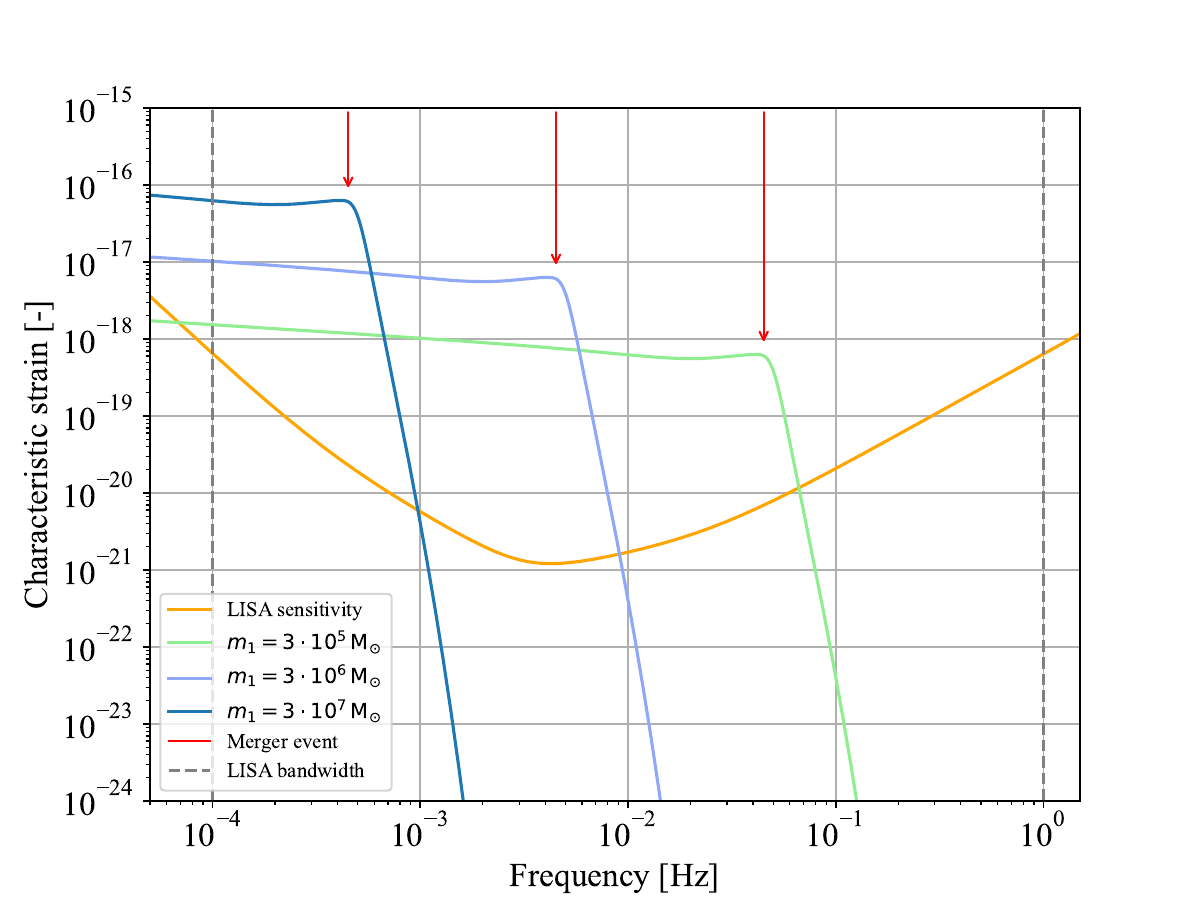}
    \caption{LISA sensitivity in the relevant frequency band, including examples of massive black hole binary mergers for $m_1=3\cdot10^5\,\text{M}_{\odot}$, $3\cdot10^6\,\text{M}_{\odot}$, and $3\cdot10^7\,\text{M}_{\odot}$, $q=m_2/m_1=1/3,\,m_1>m_2$ and further relevant parameters given in Table \ref{tab:MBHBParam}. The figure also indicates the merger event for each system. The inspiral phase lies to the left of the indication, while the behavior of the curves to the right of the merger defines the ringdown. Note that for deterministic signals, the upper representation lacks a formal definition; nevertheless, it can be approximated as explained in \cite{colpi2024lisa}. The figure is generated using portions of the code provided in \cite{LATWKatz}.}
    \label{fig:MBHBvsSens}
\end{figure}

Validation against a range of numerical relativity waveforms has shown that IMRPhenomD can model the amplitude and phase of gravitational waves across the spectrum of binary black hole configurations of interest to gravitational wave observatories like LISA. Figure \ref{fig:MBHBvsSens} depicts the LISA sensitivity, including the IMRPhenomD waveforms for three binary black hole mergers at different masses with the relevant source parameters given in Table \ref{tab:MBHBParam}. The waveforms show changes in strain amplitude and frequency of the gravitational waves throughout the inspiral, merger, and ringdown phases. The illustration assists in qualitatively assessing the detectability of such events by LISA, particularly highlighting areas where the waveforms exceed the sensitivity threshold. Lighter systems may be observable during their inspiral phase for extended periods within the measurement bandwidth, while heavier systems, although larger in amplitude, shift to lower frequencies and may remain obscured by noise for a longer pre-merger interval.

\begin{table}[h!]
\centering
\captionsetup{justification=centering}
\caption{Overview of the selected waveform parameters.}
\begin{tabular}{>{\raggedright}p{3.65cm}>{\raggedright\arraybackslash}p{3.5cm}} 
\hline\hline
\centering
\textbf{Parameter} & \textbf{Value}  \\
\hline
Component mass $m_1$  & $3\cdot10^5\,\text{M}_{\odot}$, $3\cdot10^6\,\text{M}_{\odot}$, $3\cdot10^7\,\text{M}_{\odot}$ \\
Mass ratio $q=m_2/m_1$  & $1/3$ \\
Spins $\chi_1, \chi_2$  & 0.5, 0.7 \\
Luminosity distance $D$  & 15 Gpc \\
Reference phase $\phi_{\text{ref}}$ & $0.6$ \\
\hline\hline
\end{tabular}
\label{tab:MBHBParam}
\end{table}

\subsection{LISA Response to Massive Black Hole Binaries}
To analyze how LISA responds to gravitational waves from massive black hole binaries, we begin by projecting the strain signal onto the six laser links using the \texttt{lisagwresponse} Python package \cite{https://doi.org/10.5281/zenodo.6423435}. Following this projection, the signal undergoes propagation through Time-Delay Interferometry (TDI) utilizing \texttt{pyTDI} \cite{https://doi.org/10.5281/zenodo.6351736}. 

TDI is a technique developed to mitigate laser frequency noise in post-processing \cite{Tinto2002TDI1stGen, Tinto_2004, TDITINTO2005}. Laser frequency noise arises from the inherently unstable laser sources used for the interferometry measurements, coupled with the unequal interferometer arm lengths of LISA's optical architecture. The latter introduces differential time delays in the recombined laser beams. The noise observed in the data scales proportionally with the differential delays, exceeding the gravitational wave signals by seven to eight orders of magnitude \cite{Rüdiger2008}. TDI solves this fundamental problem of spaceborne detectors like LISA by combining measurements taken along the arms of the constellation to reduce the effects of laser frequency noise. This is achieved by precisely delaying and subtracting the laser light paths to form an interferometric measurement, which cancels the laser noise common to multiple paths. The result is a set of TDI channels largely free of laser frequency noise, allowing for the detection of astrophysical signals with much greater sensitivity.

TDI  comprises different generation. TDI 1.0 targets static, non-rotating constellation geometries, while TDI 2.0 extends its noise reduction capability to account for spacecraft motion. Multiple TDI combinations have been found in the past,  each exhibiting unique responses to signal and noise \cite{muratore2020revisitation}. For example, the TDI combinations A, E, and T are synthesized orthogonal data channels that optimize gravitational wave signal detection while minimizing noise \cite{LISAOptSens}. The channels A and E are specifically attuned to gravitational waves, offering independent measurements, while T acts as a noise monitor, aiding in noise identification and mitigation.

\begin{figure}[]
    \centering
        \centering
        \includegraphics[trim=10 0 0 0, clip, width=0.47\textwidth]{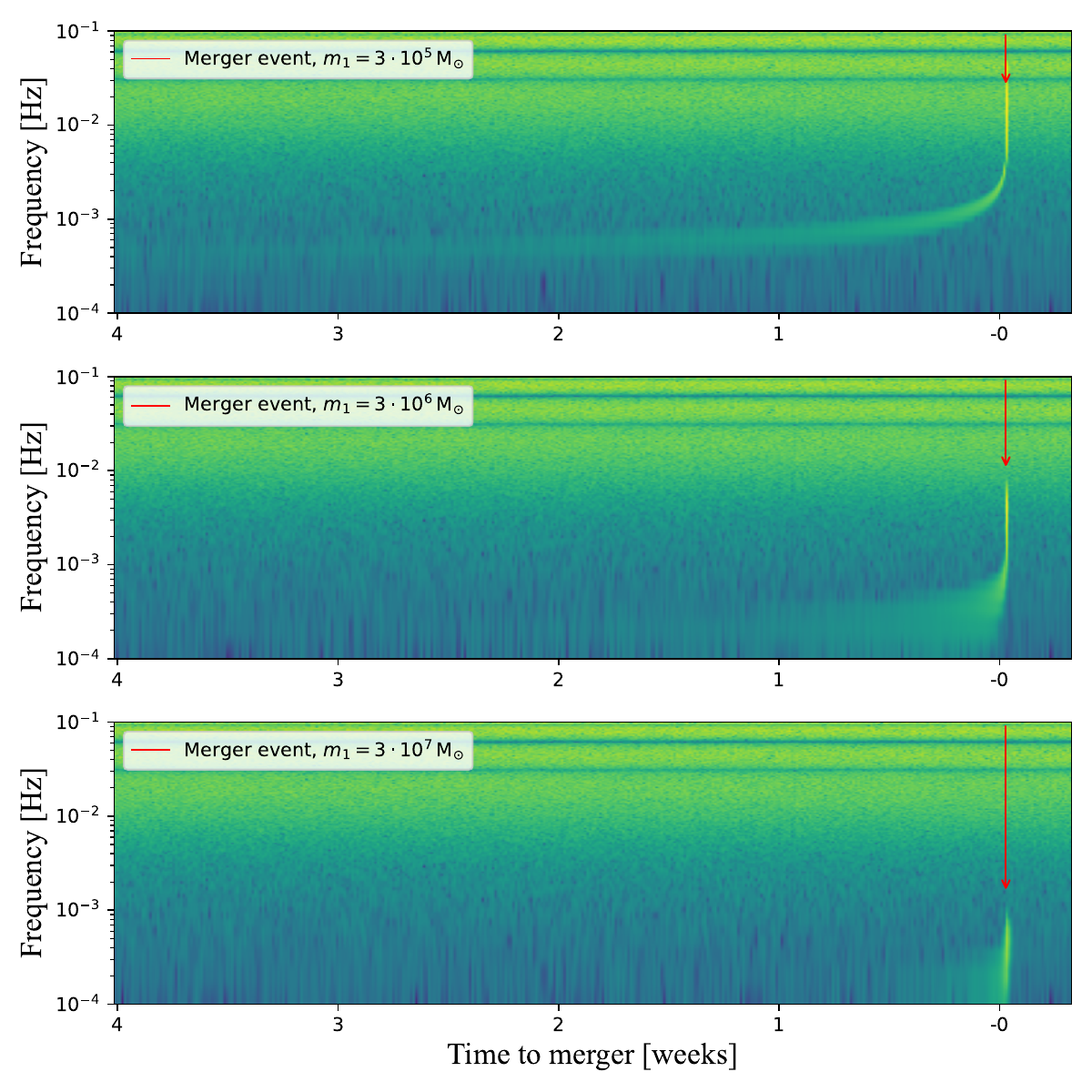}
    \caption{Spectrograms of the TDI 2.0 A channel for the three examples of massive black hole binaries depicted in Fig. \ref{fig:MBHBvsSens}. The time at which the recorded signal notably diverges from the underlying noise floor can vary significantly depending on the source parameters, the distance from the source to the constellation, and their relative orientation. This variation can range from weeks before the merger to days and hours. Note that the dark horizontal lines observed at 30 and 60 mHz represent characteristic zeros of the TDI algorithm and do not originate from astrophysical phenomena.}
    \label{fig:Specs_A123_Noisy}
\end{figure}

Figure \ref{fig:Specs_A123_Noisy} presents three spectrograms of the TDI 2.0 A channel for the massive black hole binary systems of Fig. \ref{fig:MBHBvsSens} and time-variant arm lenghts. The spectrograms illustrate how the gravitational wave signals eventually diverge from the underlying noise floor as the merger approaches. The timing of this divergence is subject to variation, influenced by the source parameters, the distance from the source to the LISA constellation, and their relative orientation. The sky localization is chosen randomly here but equal throughout the three scenarios. Instrumental noise is considered in accordance with the specifications outlined in the LISA science requirement document \cite{SciReqDoc}. The visibility of the inspiral phase ranges from weeks before the merger down to days and hours.

\noindent
In the next section, we will analyze time-evolving TDI spectrograms using a convolutional neural network to detect signals at the earliest possible stage during the inspiral. The network will be designed to detect the vertical pixel position that indicates the trace of power changes in the input data. Ground truth data is acquired from a noise-free scenario. Our training dataset will also include spectrograms without any merger, where the ground truth is designated as a negative value. Although the vertical pixel position lacks relevance in our context, we utilize it to signify a detection: a positive network output denotes detection, whereas a negative value indicates no detection.

\vspace{-10pt}
\section{Detection of Massive Binary Black Holes via Supervised Learning}\label{sec:levelDet}
The section is dedicated to detecting signals from massive black hole binary systems during their inspiral phase, marking the first step in our pipeline before predicting the future merger times of the detected signals.
Traditional anomaly detection methods utilize diverse techniques such as statistical analysis, clustering, and dimensionality reduction, each tailored to specific contexts and offering distinct advantages. The authors of this paper have previously developed a network specifically designed to identify glitches in image-transformed TDI time series data. The section will repurpose this network architecture for the specialized application outlined.
\subsection{Convolutional Neural Network Design}
Convolutional neural networks are a specialized type of feed-forward neural network. Unlike traditional fully connected feed-forward networks, where each neuron in a layer is connected to every neuron in the subsequent layer, convolutional neural networks employ convolutional layers. These layers process input data in smaller segments using cascaded convolution kernels, thus significantly reducing the number of required parameters. This allows convolutional neural networks to handle complex data like images more efficiently, as they can automatically learn and optimize filters, reducing the need for extensive preprocessing.

The research of \cite{houba2024detection} has successfully demonstrated the suitability of convolutional neural networks for detecting transient signals in LISA data, particularly in identifying glitches. Building on this, the architecture is being reapplied to the current problem dealing with the detection of signals from massive black hole binary systems within time-evolving TDI spectrograms.

\begin{figure}[]
	\centering
  \includegraphics[trim=5 20 0 0, clip,width=0.48\textwidth]{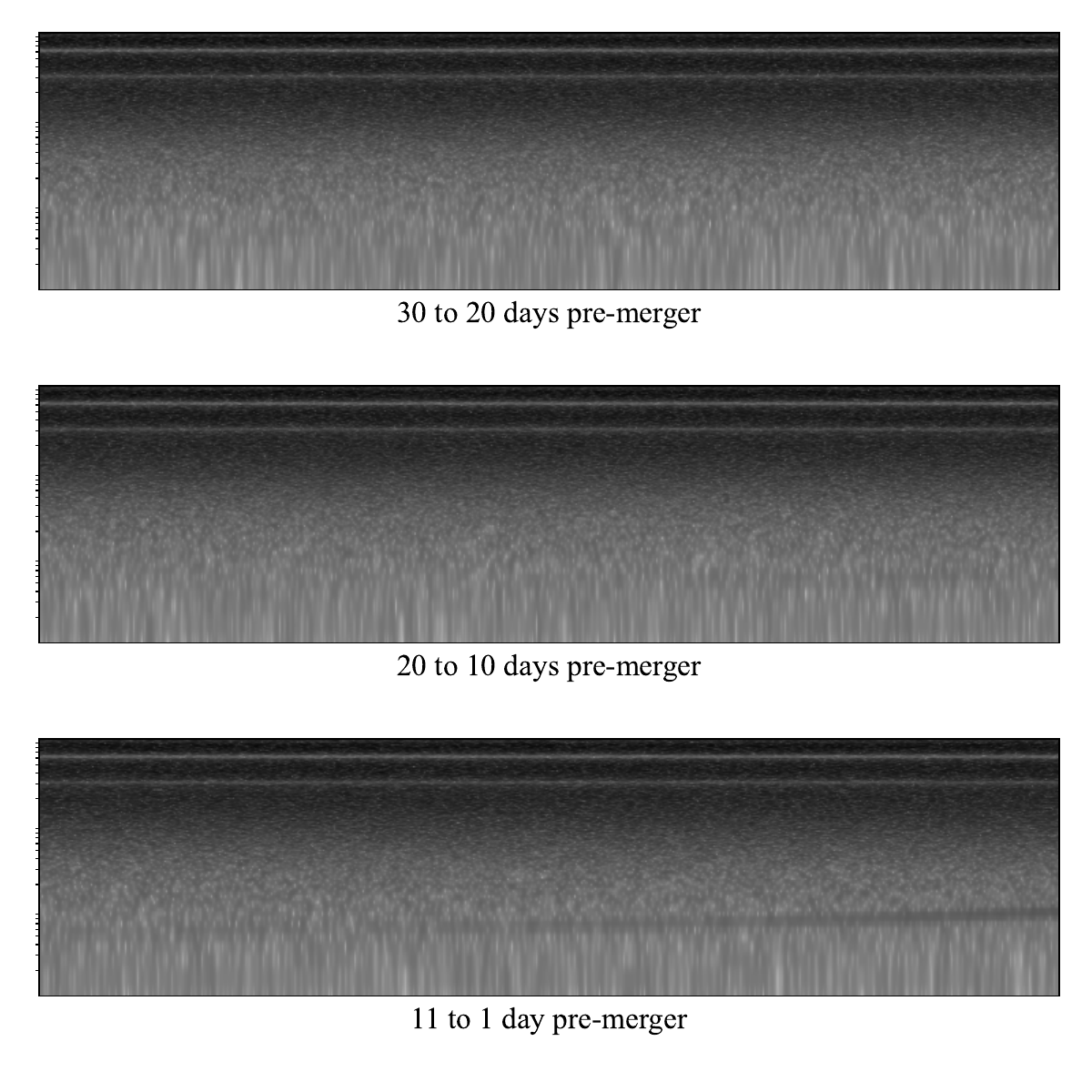}
	\caption{Examples of time-evolving black and white spectrograms of TDI 2.0 A used as network input for the purpose of signal detection. As new data points replace old ones, the spectrograms dynamically evolve. All three images are based on the merger event from Fig. \ref{fig:Specs_A123_Noisy} with  $m_1 = 3\cdot10^5\,\text{M}_{\odot}$. The upper image covers 30 to 20 days pre-merger, the middle spans 20 to 10 days pre-merger, and the lower spans 11 to 1 day pre-merger. The network aims to detect the mean vertical position of the dark trace, indicating time- and frequency-dependent changes in power during the inspiral phase. While the lower image prominently exhibits changes in signal power, such changes are barely discernible to the human eye in the middle and upper image. This motivates the investigation of convolutional neural networks in detecting subtle signal changes within LISA's noisy TDI spectrograms. Note that the spectrograms are presented to the network at a reduced resolution to expedite training and evaluation.}
	\label{im:TimeEvolvTDISpecs_Noisy}
\end{figure}

Time-evolving TDI spectrograms are derived from past measurements and updated continuously as new data becomes available. These spectrograms are useful for detecting signals from massive black hole binary systems, particularly during the inspiral phase leading up to the merger. To ensure timely detection, the analysis relies on data up to the present moment, calculated based on a selected window of past measurements. In our study, the best results were obtained with spectrograms based on 10 days of data considering the foreseen downlink sampling rate of 4 Hz. As new data points replace old ones, the spectrograms dynamically evolve. Figure \ref{im:TimeEvolvTDISpecs_Noisy} presents an example using the merger event detailed in Fig. \ref{fig:Specs_A123_Noisy} with $m_{1} = 3\cdot10^5\,\text{M}_{\odot}$. Three black and white spectrograms are plotted, each representing an example image the convolutional neural network will analyze. The upper spectrogram covers the period from 30 to 20 days pre-merger, the middle spans from 20 to 10 days pre-merger, and the lower covers the period from 11 to 1 day pre-merger. While the lower image prominently exhibit changes in signal power, such changes are barely discernible to the human eye in the middle and upper image due to the surrounding noise. The network's objective is to identify the mean vertical position of the dark trace within these spectrograms, indicating the time- and frequency-dependent changes in power associated with the inspiral phase.

\begin{figure}[]
	\centering
  \includegraphics[trim=0 280 410 0, clip,width=0.5\textwidth]{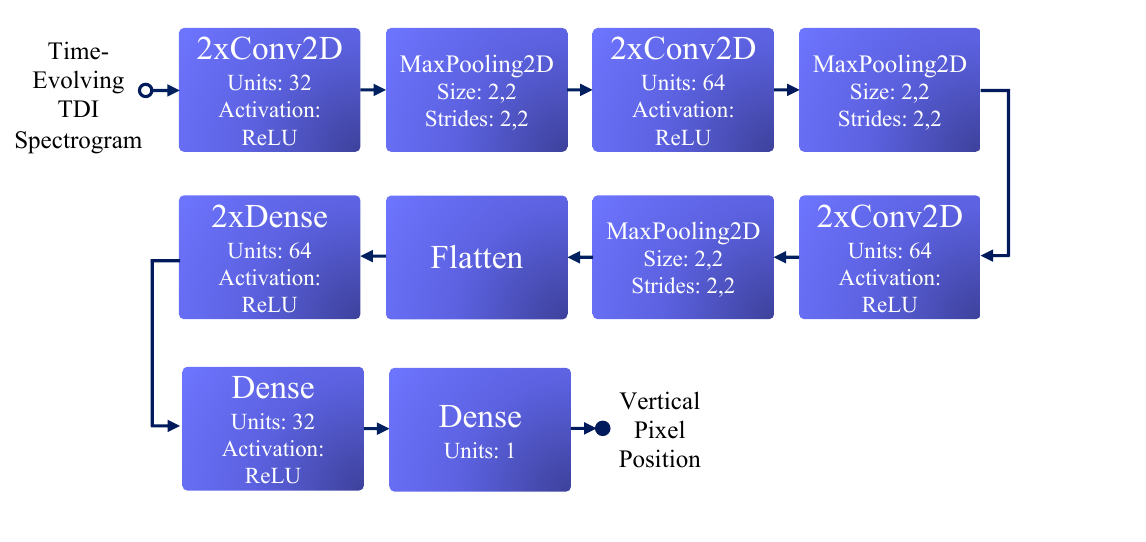}
	\caption{Architecture and attributes of the convolutional neural network employed for the detection of massive black hole binaries in time-evolving TDI spectrograms.}
	\label{im:SLNetwork}
\end{figure}

The network architecture employed for the task is depicted in Fig. \ref{im:SLNetwork}. The network begins with two convolutional layers, each featuring 32 filters of size 3×3, followed by four convolutional layers with 64 filters each, all activated by the rectified linear unit (ReLU) function to capture nonlinear relations inherent in the TDI spectrograms. Each pair of convolutional layers is succeeded by a 2×2 maximum pooling layer, which reduces spatial dimensions, computational load, and parameters. Maximum pooling layers downsample feature maps while retaining essential information, prioritizing features with higher numerical values indicative of critical information like edges or specific textures. This abstraction aids in generalizability and robustness, reducing the risk of overfitting. The network then flattens the two-dimensional feature maps into a one-dimensional vector compatible with fully connected dense layers. Three successive fully connected layers with 64, 64, and 32 nodes refine the feature representation, focusing on subtle aspects that might indicate the inspiral phase to be detected.

\subsection{Detection Performance}
We simulated 250,000 different merger events to generate the training dataset with component masses ranging from $10^5$ to $10^7$ $\text{M}_{\odot}$ and luminosity distances up to 100 Gpc, mirroring the parameters of the LISA Spritz dataset. From each merger event, we extract 100 random 10-day periods covering the interval from 1 month before the merger up to a few minutes before the merger. This results in a dataset comprising 2.5 million TDI spectrograms, incorporating both merger events and additional noise in accordance with the specifications outlined in the LISA science requirement document \cite{SciReqDoc}. The ground truth label is generated for each TDI spectrogram based on the corresponding noise-free merger scenario. The vertical position of the dark traces indicating the inspiral phase is automatically identified for each training sample in those noise-free spectrograms using Canny edge detection \cite{4767851}, thus accelerating the generation process for the training and testing datasets. To obtain a balanced training dataset, we augmented it with an additional 2.5 million merger-free TDI spectrograms. These spectrograms are labeled with a negative value to denote the absence of a signal.
A detection is defined based on the neural network's output: a positive output for a TDI input spectrogram indicates a detection, while a negative output signifies no detection. Additionally, the time to the merger is saved for every testing sample to enable plotting the rate of correct detections against the time to merger.

The testing set comprises 20,000 merger events characterized by the source parameters outlined in Table \ref{tab:MBHBParam}, additional 20,000 merger events featuring a variation in luminosity distance from 15 Gpc to 20 Gpc, and 20,000 merger-free samples. 

\begin{figure}[]
	\centering
  \includegraphics[trim=5 20 0 0, clip,width=0.48\textwidth]{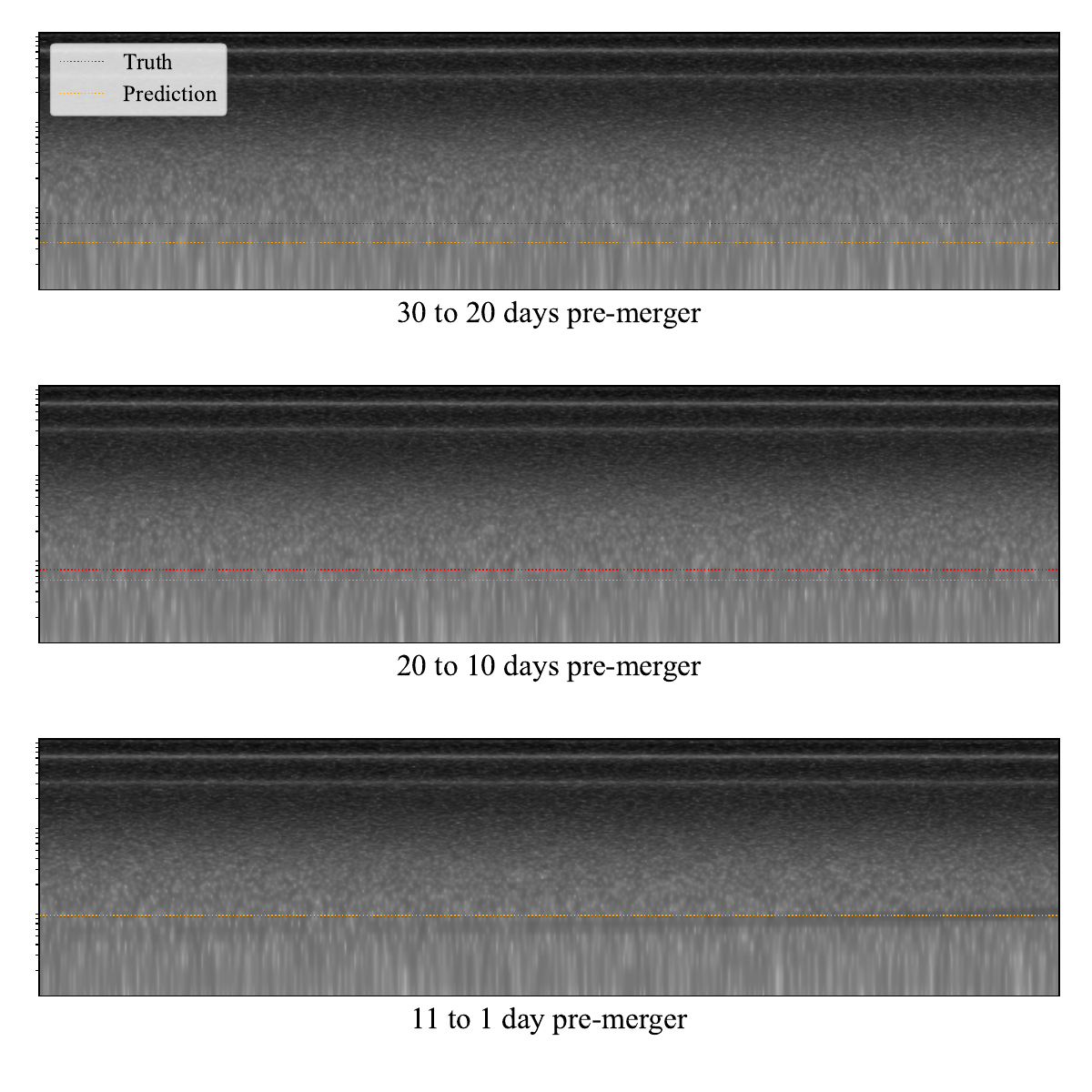}
	\caption{Network prediction for the example of Fig. \ref{im:TimeEvolvTDISpecs_Noisy}. The network correctly identifies a signal in all three TDI spectrograms.}
	\label{im:TimeEvolvTDISpecs_Noisy_NNOutput}
\vspace{15pt}
	\centering
  \includegraphics[trim=5 0 0 0, clip,width=0.47\textwidth]{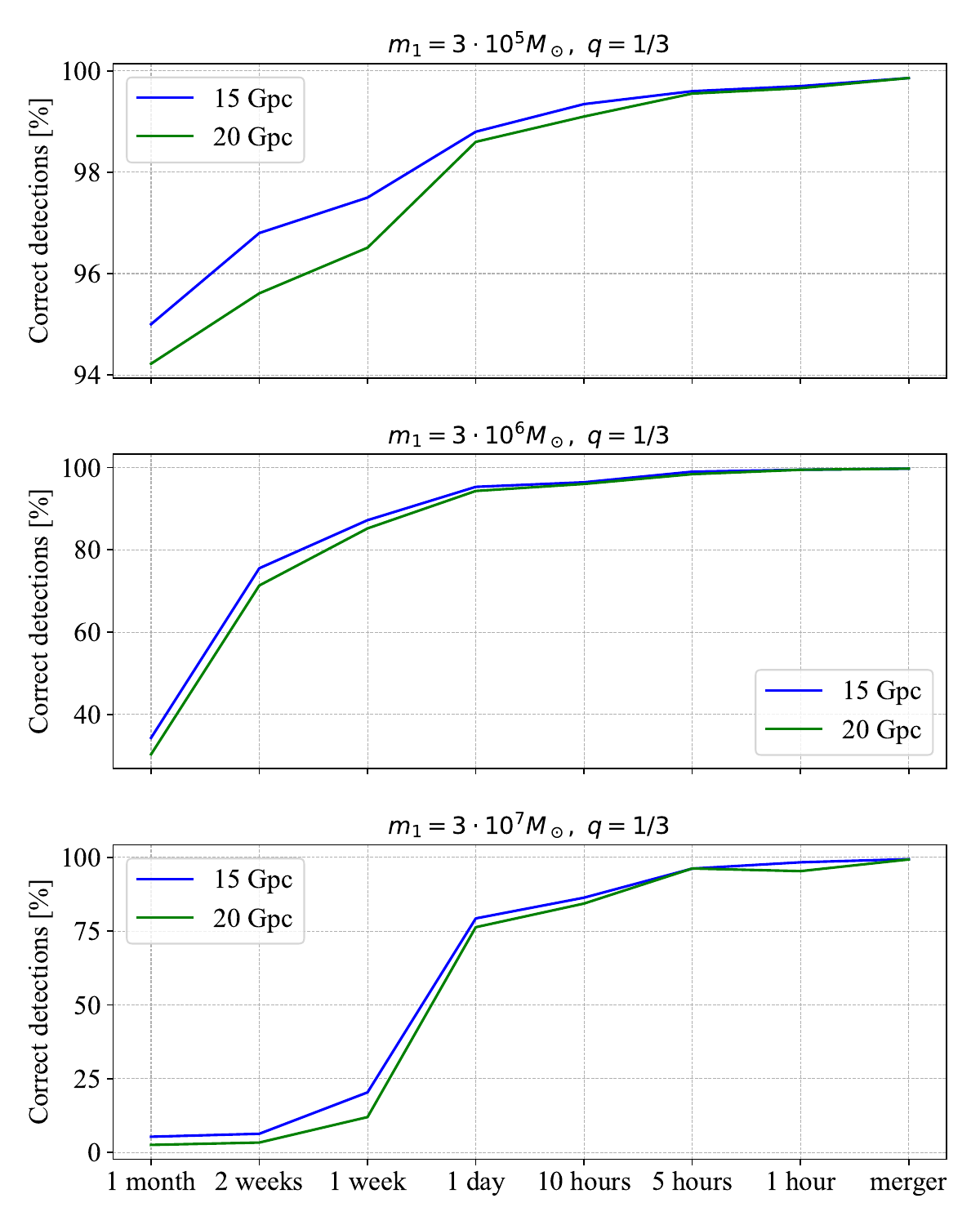}
	\caption{Signal detection rate for the three merger scenarios depicted in Fig. \ref{fig:Specs_A123_Noisy} and two different luminosity distances.}
	\label{im:DetRates_CNN}
\end{figure}

The network's prediction for the example depicted in Fig. \ref{im:TimeEvolvTDISpecs_Noisy} is presented in Fig. \ref{im:TimeEvolvTDISpecs_Noisy_NNOutput}. Note that the primary objective is not to precisely replicate the ground truth. Instead, the goal is for the network's output to be positive if and only if TDI spectrograms contain a recorded signal from a massive black hole binary, thus triggering detection.

The performance of signal detection for the three scenarios depicted in Fig. \ref{im:TimeEvolvTDISpecs_Noisy}, along with two different luminosity distances, is illustrated in Fig. \ref{im:DetRates_CNN} for 1 month prior to the merger, analogous to \cite{Fabian2019AthenaLISAS}. Consistent with Figs. \ref{fig:MBHBvsSens} and \ref{fig:Specs_A123_Noisy}, we observe that as the total mass increases, early signal detection becomes more challenging. 

In this initial design, we focus solely on utilizing the TDI 2.0 A channel. However, there is potential for future enhancements by considering modifications to the network architecture. One such possibility is adjusting the network to accommodate two or three TDI spectrograms simultaneously as input, which may further improve the detection performance. Given the large multidimensional parameter space involved, it is important to acknowledge that more evaluations would be necessary to fully demonstrate the network's detection capabilities. While the results presented may not be entirely surprising and have been addressed by other methodologies in the past, it should be emphasized that this network serves as the first component within the machine-learning-based low-latency alert prototype pipeline presented in this paper. 

The second step focuses on predicting the future merger time in instances where the convolutional neural network has identified a signal.  The task presents a high level of complexity. For this, the suitability of reinforcement learning will be explored in the following.

\section{Reinforcement Learning in Prediction Tasks}\label{sec:level2}
This section provides an overview of the theoretical foundations of reinforcement learning, particularly focusing on the design of a reinforcement learning framework for predicting the timing of future massive black hole mergers observed by LISA. It aims to illustrate how the core concepts and methodologies of reinforcement learning theory can be leveraged to enhance the precision and reliability of such predictions.

\subsection{Introduction to Reinforcement Learning}

Reinforcement learning is a subset of machine learning focused on enabling an autonomous agent to learn optimal decision-making through direct interaction with its environment. The interaction is driven by trial and error, where the agent performs actions and receives feedback in the form of rewards, contrasting sharply with supervised learning paradigms that rely on pre-labeled datasets for model training. In reinforcement learning, the agent's learning is self-guided by its experiences, with the overarching goal to maximize the accumulation of rewards over time, a process that involves learning to predict the long-term consequences of actions taken in various states of the environment. More details can be found in \cite{Sutton2018-md, 8714026,Fran_ois_Lavet_2018, omi2023dynamic, vajjha2020certrl}.

Central to reinforcement learning is the mathematical formalism of Markov decision processes (MDPs) \cite{LITTMAN20019240, Buerle2011, 7260117}, a framework that models sequential decision-making problems where outcomes are partly controlled by a decision-maker (the agent) and partly random. An MDP is formally defined by the following components:
\vspace{15pt}\newline\noindent\
\textbf{States} \(s \in S\): A finite or infinite set of all possible situations or configurations the environment can be in. Each state encapsulates all the necessary information that the agent needs to make a decision at any given point in time.
\\
\vspace{0pt}\newline\noindent\textbf{Actions} \(a \in A\): For each state, there is a set of actions that the agent can choose from. Actions are the mechanisms through which the agent interacts with the environment, transitioning from one state to another. The set of actions available to the agent can be either discrete or continuous. In discrete action spaces, the agent oftentimes chooses from a finite set of distinct actions, making the selection process straightforward but potentially limiting the agent's control granularity. Conversely, continuous action spaces allow the agent to select actions from a continuous range, offering a finer degree of control at the cost of more complex decision-making and optimization challenges. Note that in our context, actions denote time-to-merger predictions.
\\
\vspace{0pt}\newline\noindent\textbf{Transition Probability} \(P\):  The transition probability function \(P: S \times A \rightarrow S\) defines the dynamics of the environment. For each state \(s\) and action \(a\), \(P(s' | s, a)\) gives the conditional probability of transitioning to a new state \(s'\) after taking action \(a\) in state \(s\).
\\
\vspace{0pt}\newline\noindent\textbf{Reward} \(R\): The reward function \(R: S \times A \times S \rightarrow \mathbb{R}\) assigns a numerical reward to each transition between states. The reward received after executing action \(a\) in state \(s\) and moving to state \(s' \) is denoted by \(R(s, a, s')\) or briefly $R$, representing the immediate payoff of that action.
\\
\vspace{0pt}\newline\noindent\textbf{Discount Factor} \(\gamma\): In an MDP, the discount factor $\gamma$ balances the importance of immediate rewards versus delayed rewards. It is represented by: 
\begin{equation}
R_t = \sum_{k=0}^\infty \gamma^k \cdot R_{t+k+1},  \label{eq:Rtgamma}   
\end{equation}
where $R_t$ denotes the total discounted reward starting from time step $t$, $R_{t+k+1}$ is the reward obtained at time step $t+k+1$, $\gamma \in [0,1)$ is the discount factor, and $k$ represents the time step difference from the current time $t$. The term $\gamma^k$ discounts future rewards, with lower $\gamma$ values emphasizing immediate rewards and higher $\gamma$ values prioritizing future rewards. Thus, $\gamma$ influences the agent's decision-making by determining how it weighs the trade-off between immediate and future rewards.
\newline\newline\noindent
The objective of a reinforcement learning agent within the MDP framework is to discover a policy \(\pi: S \rightarrow A\), which is a mapping from states to actions, that maximizes the expected cumulative reward, often expressed as the expected sum of discounted rewards, see Eq. \eqref{eq:Rtgamma}. The policy essentially represents the agent's strategy or plan of action for any given state, formulated with the intention of maximizing rewards.

A critical challenge in reinforcement learning is the exploration-exploitation dilemma, which involves making strategic decisions about when to explore new, potentially better strategies (exploration) and when to utilize existing knowledge to achieve the best immediate outcome (exploitation). Finding an optimal balance between these two approaches is crucial for the effective learning and performance of a reinforcement learning agent. Exploration enables the agent to collect new information that could lead to better long-term outcomes, while exploitation allows the agent to rely on its current knowledge to maximize immediate rewards. The efficacy of reinforcement learning algorithms hinges on their ability to navigate this dilemma, adapting their strategy as they learn more about the environment to improve their decision-making process progressively \cite{Wilson2021, BergerTal2014}.

 \begin{figure}[b!]
	\centering
  \includegraphics[trim=15 250 410 50, clip,width=0.5\textwidth]{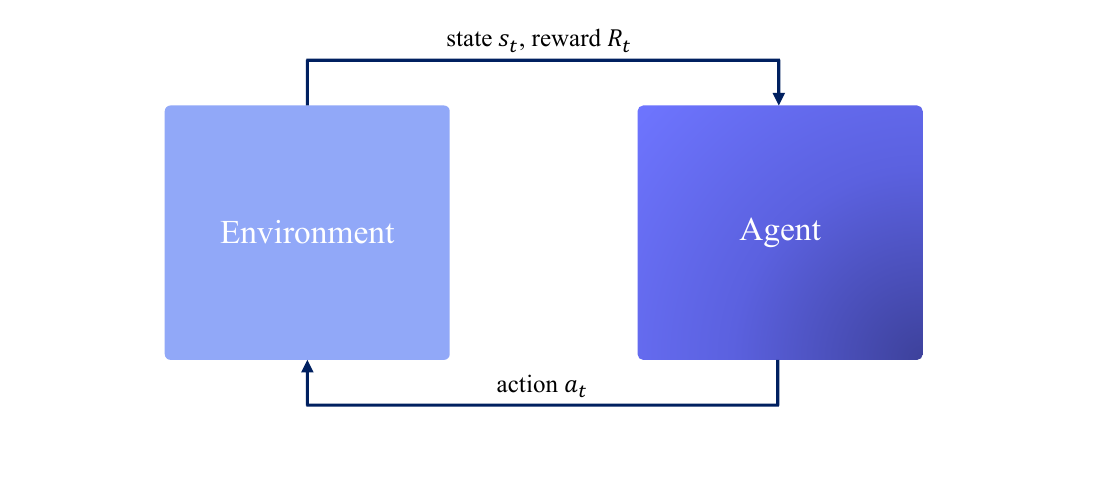}
	\caption{Concept of reinforcement learning. The figure shows the interaction between the environment and the agent. The environment represents the external system or context within which the agent operates, while the agent embodies the decision-making entity that interacts with and learns from the environment.}
	\label{im:RLBasicSetup}
\end{figure}
The basic principle of reinforcement learning is illustrated in Fig \ref{im:RLBasicSetup}. In summary, reinforcement learning offers a powerful framework for autonomous decision-making in uncertain environments underpinned by the mathematical structure of MDPs. Through exploration and exploitation, guided by the feedback loop of actions and rewards, reinforcement learning agents learn to devise strategies that optimize their performance according to a well-defined notion of cumulative reward, demonstrating broad applicability across various domains requiring adaptive behavior. 
The predictive capability of reinforcement learning shall be studied in the context of merging massive black hole binaries observed by LISA. The task of accurately predicting such complex and dynamic events demands a deep understanding of the temporal relationships between observations, something that reinforcement learning is expected to handle. Before delving into the specific reinforcement learning framework designed for this task, the concepts of temporal difference learning and Q-learning are introduced. These methodologies are critical for developing algorithms that can learn from and adapt to the evolving patterns observed in LISA's TDI measurements.

\subsection{The Concepts of Temporal Difference Learning and Q-Learning}
Temporal difference learning represents a methodology within the reinforcement learning spectrum that enables reinforcement learning agents to deduce optimal policies through iterative interactions with their environment \cite{Watkins1992, Uther2011, Jang2019}. In reinforcement learning, a policy is a strategy employed by the agent, defined as a mapping from states of the environment to actions the agent should take when in those states. Essentially, a policy \(\pi\) dictates the course of action an agent follows, aiming to maximize some measure of long-term reward. Optimal policies are those that yield the highest expected reward from any given state, guiding the agent to its goals most efficiently.

Temporal difference learning and Q-learning, a specific temporal difference algorithm, merge the advantages of Monte Carlo methods and dynamic programming, facilitating a pathway for agents to learn and adapt directly from raw experience. Unlike Monte Carlo methods, which require the completion of episodes for updates, temporal difference learning allows agents to learn from incomplete sequences, adjusting their policy after each step based on a combination of observed rewards and estimates of future rewards. This capability makes temporal difference learning particularly suited for continuous tasks without defined endpoints. Furthermore, temporal difference learning circumvents the need for a model of the environment's dynamics, a requirement in traditional dynamic programming, thus enabling agents to operate and learn in unknown or complex settings.

In this study, we employ Q-learning to develop an optimal policy that shall empower an agent to accurately forecast the future merger timelines of massive black hole binaries. Q-learning focuses on learning the quality value, short Q-value, of taking specific actions in specific states, thus learning the optimal policy indirectly. It updates the action-value function, also known as the Q-function, which estimates the expected quality of taking a given action in a given state and following the optimal policy. Through iterative adjustments of the Q-values, guided by the action-value function, Q-learning allows the agent to move towards an optimal policy, even when lacking knowledge of the environment's model.

\subsubsection{Action-Value Function}
Central to the discussion of Q-learning is the concept of the action-value function, commonly denoted as \(Q^\pi(s_t, a_t)\). The action-value function represents the expected return, i.e., the cumulative discounted future rewards, of taking an action \(a_t\) in a state \(s_t\) at time step $t$ and following a policy \(\pi\). Mathematically, the action-value function for a policy \(\pi\) is expressed as:

\begin{equation}
Q^\pi(s_t, a_t) = \mathbb{E}_\pi\left[\sum_{k=0}^{\infty} \gamma^k R_{t+k+1} \,|\, S_t = s_t,\, A_t = a_t\right],
\end{equation}
where \(\gamma\) is the discount factor reflecting the importance of future rewards, \(R_{t+k+1}\) is the reward received \(k\) steps in the future, and the expectation \(\mathbb{E}_\pi\), which is taken over the distribution of possible paths that follow from taking action \(a_t\) in state \(s_t\) under policy \(\pi\) at time step \(t\). Then, \(Q^\pi(s_t, a_t)\) estimates the expected return of taking action \(a_t\) in state \(s_t\) and following policy \(\pi\).

Q-learning specifically targets the optimal action-value function, denoted as \(Q^*(s_t, a_t)\), which represents the expected return of taking action \(a_t\) in state \(s_t\) and following the optimal policy. The essence of Q-learning lies in iteratively updating the Q-values towards this optimal function using the experiences sampled from the environment, according to the update rule:

\begin{align}
\begin{aligned}
        &Q^\pi(s_t, a_t) \leftarrow Q^\pi(s_t, a_t) \\ &+  \alpha \left[R_{t+1} + \gamma \max_{a'} Q^\pi(s_{t+1}, a') - Q^\pi(s_t, a_t)\right].
\end{aligned}\label{eq:QLUpdateRule}
\end{align}
Here, \(\alpha\) denotes the learning rate, controlling the extent to which new information overrides old information, and  $\max_{a'} Q^\pi(s_{t+1}, a')$ represents the value of the best possible action from all possible actions  $a'$ at the next state \(s_{t+1}\), encapsulating the essence of policy improvement in dynamic programming.

\subsubsection{The Bellman Equation}

The theoretical foundation of Q-learning and the broader field of reinforcement learning rests upon the Bellman equation, which provides a recursive decomposition of the action-value function \cite{Chin2021, HarvardWebsite1}. Specifically, for the optimal action-value function \(Q^*(s_t, a_t)\), the Bellman optimality equation is given by:

\begin{align}
\begin{aligned}
&Q^*(s_t, a_t) = \\  &\mathbb{E}\left[R_{t+1} + \gamma \max_{a'} Q^*(s_{t+1}, a') \,|\, S_t = s_t,\, A_t = a_t\right].
\end{aligned}
\end{align}
This equation asserts that the value of taking action \(a_t\) in state \(s_t\) is equal to the expected immediate reward \(R_{t+1}\) plus the discounted value of the state reached thereafter, \(s_{t+1}\), assuming optimal behavior henceforth. The recursive nature of the Bellman equation enables the iterative approximation of \(Q^*(s_t, a_t)\) through the Q-learning update rule of Eq. \eqref{eq:QLUpdateRule}, providing a powerful mechanism for policy evaluation and improvement without the need for a model of the environment's dynamics.

\noindent
In summary, Q-learning furnish reinforcement learning agents with the mechanisms to infer optimal policies via direct interaction with their environment, harnessing the recursive structure of the Bellman equation to update their knowledge base in the form of Q-values based on real experiences. 
\vspace{-7pt}
\subsubsection{From Traditional Q-Learning to Deep Q-Learning}
Deep Q-learning extends traditional Q-learning by incorporating deep neural networks as function approximators for the action-value function \(Q^\pi(s_t, a_t)\). This enables the handling of high-dimensional state spaces typical in complex environments, where traditional Q-learning with tabular methods would be impractical due to the curse of dimensionality. Tabular methods maintain a lookup table that stores the value of the action-value function for every possible state-action pair in the environment. This approach is straightforward and effective in environments with a relatively small number of discrete states and actions, however, they encounter significant challenges in more complex scenarios. The primary limitation is the scalability issue: as the number of states or actions increases, the size of the Q-table can grow exponentially \cite{cummins2024reinforcement}. This demands extensive memory resources and requires the agent to visit each state-action pair multiple times to learn accurate values, which becomes impractical or even impossible in environments with a large or continuous state space.

\noindent
Deep Q-learning addresses these limitations by replacing the Q-table with a deep neural network that approximates the action-value function \(Q^\pi(s_t, a_t;\,\theta)\), where \(\theta\) represents the weights of the network \cite{9681985, Tsantekidis2022}. This approach enables the handling of high-dimensional state spaces characteristic of complex or continuous environments. By learning to generalize across similar states, the neural network can estimate action values even for states it has never encountered before, significantly enhancing the agent's learning efficiency and scalability.

The traditional tabular Q-learning methods, while foundational, fall short in tackling the vast and complex data generated in this context. In applying deep Q-learning to the context of LISA observations, the goal is to develop a reinforcement learning agent that can navigate the state space of LISA data, making accurate predictions about future merger events based on accumulating observational measurements over time. 
The introduction of deep Q-learning marks an important step in applying reinforcement learning to predict future massive black hole mergers observed by LISA.
\vspace{-5pt}
\section{Reinforcement Learning Framework for Time-to-Merger Predictions in LISA}\label{sec:RLFL}

To develop a reinforcement learning framework dedicated to predicting the time-to-merger of massive black holes, as observed by LISA, a sequence of systematic steps is presented and detailed subsequently. Following these steps was found to be helpful for guiding the structured development and implementation of the reinforcement learning model, ensuring its adequacy for addressing the complexity inherent to this predictive task.
\\
\vspace{5pt}\newline\noindent
{1. Definition of the environment and state design:} In the proposed framework, states are constructed to reflect the temporal evolution of gravitational wave signals leading to a merger, drawing upon data from LISA's TDI channels. The pre-processing of TDI channel data into a structured representation suitable for state definition is a critical component of the environmental model. This pre-processing involves storing past observations up to a predetermined extent, which are then provided to the agent. This incorporation of past data aims to improve the agent's ability to make informed decisions by recognizing patterns that may indicate imminent mergers.
The environment definition allows various approaches to predict the time until a merger. These include predictions relative to the current time step, predictions of the absolute merger time, and predictions of the merger time relative to the previous event. The latter two approaches are generally considered less favorable. Predicting the absolute merger time requires the model to output progressively larger values, potentially impacting the stability of the model. Additionally, basing predictions on the time since the last merger can lead to cumulative inaccuracies if previous predictions contained errors. The time-to-merger predictions discussed in this paper are stated in the LISA frame described in \cite{PhysRevD.103.083011}.
\\
\vspace{5pt}\newline\noindent
{2. Definition of the actions:} Actions refer to the predictions or decisions rendered by the reinforcement learning agent contingent upon the present state. Given the adoption of deep Q-learning in our framework, we have chosen to implement a continuous action space. This decision accommodates the nuanced and varied nature of the predictions required, facilitating a more granular and accurate estimation process. The definition of the action space aims to find an optimal balance, ensuring it encompasses a sufficiently broad range of potential decisions to address the task's complexity while remaining computationally manageable and interpretable. 
\\
\vspace{5pt}\newline\noindent
{3. Constructions of the reward function:} The formulation of the reward function is critical, as it directs the learning trajectory by offering feedback on the agent's actions. In the context of merger time predictions, the reward can be chosen to be inversely proportional to the prediction error, awarding higher rewards for more precise forecasts. An ancillary consideration might involve incorporating extra rewards or penalties tied to the proximity to the merger event, thereby fostering improved prediction accuracy as events approach.
\vspace{5pt}\newline\noindent
{4. Design of the deep neural network model for Q-function approximation:} Given the anticipated complexity and vast dimensionality of the state space, deep neural networks are typically harnessed to approximate the Q-function. The architecture of this neural network necessitates thorough design to aptly process the unique attributes of gravitational wave data and meet the specific demands of the task, encompassing considerations such as the arrangement of layers, neurons, activation functions, and learning parameters.
\\
\vspace{5pt}\newline\noindent
{5. Definition of an exploration-exploitation strategy:} Defining how the agent will navigate the trade-off between exploring the state space to uncover new knowledge and exploiting extant information to optimize outcomes is important for proficient learning. Strategies like the $\epsilon$-greedy approach or more elaborate methods like the Upper Confidence Bound or Thompson sampling can be adopted \cite{EpsilonGreedy, PeterAuer, phdthesisAbeille}.
\\
\vspace{5pt}\newline\noindent
{6. Training and evaluation protocol:} After implementing the environment model, the agent, and their interaction, instituting a stringent protocol for training the framework denotes the next step. This includes determining the number of episodes, steps per episode, and evaluative metrics to monitor progress and ensure the model's performance aligns with the set of objectives. This protocol should also integrate validation mechanisms to evaluate the model's capacity for generalization and avert overfitting. In this phase, it is crucial to determine the optimal reinforcement learning hyperparameters, including the discount factor, learning rate, and the rate at which the exploration probability decreases, to tune the learning process for optimal performance. Expanding upon this framework, we employ a multi-agent approach where multiple agents with different hyperparameters interact independently with the same environment. Following their interactions, their individual performances are compared, and the parameters are selected based on predefined criteria. This approach allows for a diverse exploration of the solution space and facilitates the discovery of robust policies that generalize well across different scenarios.
\\
\vspace{5pt}\newline\noindent
{7. Testing with simulated or real-world observational data:} The final step entails the integration of the trained reinforcement agent into a testing environment characterized by training-independent simulated or observed gravitational wave data. The steps outlined are iterative, allowing for repeated refinement based on the performance achieved during initial setup and subsequent testing.

\subsection{Framework Design and Architectural Breakdown }
Following the outlined steps, we commence by introducing the foundational elements of the reinforcement learning framework: the environment and agent classes. These core components are detailed through pseudocode, offering a comprehensive understanding of their respective functionalities and their interaction.
\subsubsection{Environment Class }
The \texttt{Environment} class is designed to encapsulate the environmental aspects of the reinforcement learning framework. The high-level architecture is presented in the following.
\vspace{10pt}
\FloatBarrier
\begin{algorithm}
\caption*{Class: \texttt{Environment}}\label{alg:Environment_class}
\begin{algorithmic}[1]
\Function{Initialize}{TDI data, merger times, params}
    \State  Initialize environment with data and parameters.
    \State {Define action and observation spaces}.
    \State {Initialize internal variables}.
\EndFunction
\Function{Step}{action}
    \State {Call \textsc{CalculateReward}.}
    \State {Call \textsc{GetState} to update internal state.}
    \State  \hspace{10pt} \textbf{return} new state and reward
\EndFunction
\Function{CalculateReward}{action}
    \State {Compute prediction error.} 
    \State  {Calculate reward based on prediction error.} 
    \State \hspace{10pt} \textbf{return} calculated reward 
\EndFunction
\Function{GetState}{step}
        \State {Obtain past TDI data up to current time step.}
        \State {Generate state vector from that.}
\EndFunction
\Function{Reset}{}
    \State {Prepare environment for a new episode.} 
    \State {Select random start point within data bounds.} 
    \State {Reset internal counters and states.} 
    \State \hspace{10pt}  \textbf{return} reset environment 
\EndFunction
\end{algorithmic}
\end{algorithm}
\noindent 1. \textsc{Initialize}: The environment initialization configures the parameters and data structures, thereby providing the simulated LISA TDI observations in an appropriately structured format. It defines the state and action spaces, accommodating the continuous nature of the actions that correspond to merger time predictions.
\\ \newline\noindent 2. \textsc{Step}: When the agent proposes an action, the environment undergoes a state update and computes the corresponding reward. These intertwined operations are integrated within the \textsc{Step} function.
\\
\vspace{5pt}\newline\noindent  3. \textsc{CalculateReward}: The reward function employed is designed to quantify the accuracy of the agent's predictions relative to actual merger events. It is formulated as follows:

\begin{align}
\begin{aligned}
R_t = R_{\text{max}}\cdot e^{-w \cdot |t_{\text{predicted}} - t_{\text{actual}}|},
\end{aligned}\label{eq:reward}
\end{align}

where:
\begin{itemize}
    \item $t_{\text{predicted}}$ is the time step at which the agent predicts the merger will occur, calculated as $t + a_t$ with $a_t$ being the action taken at time step $t$.
    \item $t_{\text{actual}}$ represents the actual time of the merger event. Note that $t_{\text{actual}}$ is inaccessible outside the training phase. Then, $t_{\text{actual}}$ is substituted by $t_{\text{TM}}$ for the purpose of testing and real-world application. Here, $t_{\text{TM}}$ is the estimated merger time used for the reward calculation as obtained by truncated waveform template matching, a methodology detailed in Section \ref{sec:level3}. This approach allows for an approximation of merger timings, facilitating the assessment and adjustment of the agent's predictions. The uncertainty in $t_{\text{TM}}$ magnifies as the merger event stretches further into the future. The reinforcement learning setup needs to deal with these uncertainties.    
    \item $R_{\text{max}}$ and $w$ are parameters that determine the magnitude of the reward and its sensitivity to the difference between $t_{\text{predicted}}$ and $t_{\text{actual}}$ or $t_{\text{TM}}$. These parameters are determined through multi-agent-based hyperparameter optimization.
\end{itemize}
\vspace{5pt}\noindent 4. \textsc{GetState}: The state at any given time step is constructed from current and past data, facilitating pattern recognition for future event predictions. The state \(s_t\) at time step \(t\) is a composite of observations over a defined window:
    \begin{equation}
    s_t = \{d_{t-h+1}, d_{t-h+2}, \ldots, d_t\},
    \end{equation}
where \(d_t\) denotes data at time step \(t\) and \(h\) represents the history size. The history size is determined again through multi-agent-based hyperparameter optimization. We use one agent for each of the three optimal TDI channels A, E, and T, resulting in three separate predictions at every time step. These predictions are then averaged to produce the final prediction of the reinforcement learning framework.
\\

\vspace{5pt}\noindent 5. \textsc{Reset}: This function reinitializes the environment for a new episode, selecting a fresh starting point within the data to ensure varied learning experiences.
\vspace{-2pt}
\subsubsection{ Agent Class} \noindent The structure of the \texttt{Agent} class is: 

\begin{algorithm}
\caption*{Class: \texttt{Agent}}\label{alg:agent_class}
\begin{algorithmic}[1]
\Function{Initialize}{params}
    \State {Call \textsc{BuildActorModel}.}
      \State {Call \textsc{BuildCriticModel}.}
    \State {Initialize target models.}
     \State {Set learning rate, exploration rate and discount factor.}
\EndFunction
\Function{BuildActorModel}{}
    \State {Defines the actor model architecture.}

\EndFunction
\Function{BuildCriticModel}{}

    \State {Defines the critic model architecture.}
\EndFunction
\Function{Act}{state}
    \State {Decide for exploration or exploitation.}
    \State {Perform an action based on current state.}
\EndFunction
\Function{Train}{state, action, reward, next state}
    \State {Predict next action using target actor model.}
    \State {Calculate future Q-value from target critic model.}
    \State {Compute target Q-value incorporating discount factor \text{\,\,\,\,\,\,\,\,\,}and reward.}
    \State {Obtain current Q-value from critic model.}
    \State {Calculate critic and actor losses.}
    \State {Compute gradients for both actor and critic models.}
    \State {Apply gradient clipping to stabilize training.}
    \State {Update actor and critic models with computed \text{\,\,\,\,\,\,\,\,\,}gradients.}

\EndFunction
\Function{UpdateTargetNetworks}{}
    \State {Update target actor and target critic models.}
\EndFunction
\end{algorithmic}
\end{algorithm}
\noindent 
Central to the \texttt{Agent} class are the actor-critic mechanism and target networks, which draw on foundational concepts in reinforcement learning and Q-learning. These elements are integrated to enhance the agent's decision-making accuracy and the stability of its learning process. The concepts will be explained in the following.
\newline
\paragraph{Actor-Critic Mechanism}
\,\\ \,\\
The actor-critic architecture is a hybrid approach that combines policy-based and value-based reinforcement learning methods. It consists of two main components: the actor, which selects actions, and the critic, which agent-internally evaluates the actions taken by the actor \cite{Abdalla2023, Parisi2019}. This mechanism is grounded again in the framework of MDPs and the Bellman equation.

\begin{figure}[b]
	\centering
  \includegraphics[trim=0 385 410 0, clip,width=0.5\textwidth]{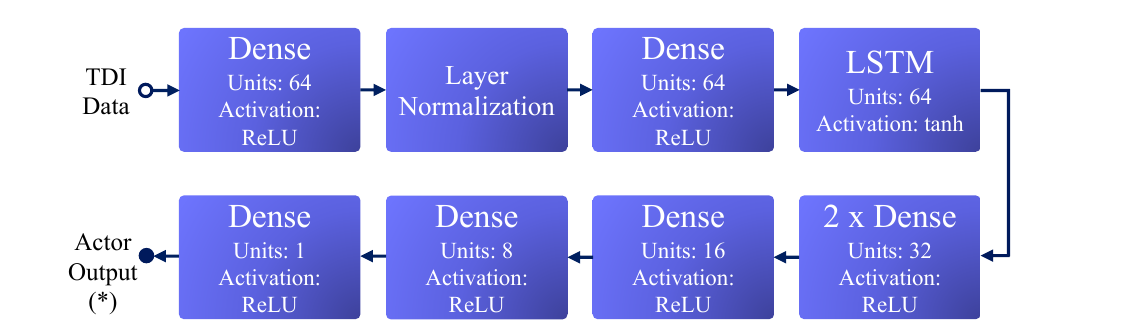}
	\caption{Architecture and attributes of the actor.}
	\label{im:NetworksAct}
\vspace{14pt}
	\centering
  \includegraphics[trim=0 217 410 170, clip,width=0.5\textwidth]{Networks.pdf}
	\caption{Architecture and attributes of the critic.}
	\label{im:NetworksCrit}
\end{figure}

\begin{enumerate}[leftmargin=*]
    \item \textbf{Actor Component:} The actor defines a policy, $\pi(a_t|s_t;\theta^\pi)$, that maps states to a distribution over actions, where $\theta^\pi$ are the parameters of the policy network. The actor's goal is to select actions that maximize the expected return from any given state. The action-value function, $Q^\pi(s_t,a_t)$, plays a crucial role here, estimating the expected return of taking an action $a_t$ in state $s_t$ under policy $\pi$. Mathematically, the actor updates its policy parameters $\theta^\pi$ in the direction that maximizes the expected return by gradient ascent on the policy's performance objective:
    \begin{equation}
    \theta^\pi_{t+1} \leftarrow \theta^\pi_t + \alpha \nabla_{\theta^\pi} \log \pi(a_t|s_t;\theta^\pi) Q^\pi(s_t,a_t),
    \end{equation}
    where $\alpha$ is the learning rate for the actor.
    
    \item \textbf{Critic Component:} The critic estimates the action-value function $Q^\pi(s_t,a_t;\,\theta^Q)$, with $\theta^Q$ representing the parameters of the critic network. This estimation provides a measure of the expected return of taking action $a_t$ in state $s_t$, serving as feedback for the actor's performance. The critic updates its value function based on the temporal difference error $\delta_t$, a measure of the discrepancy between the expected and actual returns:
    \begin{align}
    \begin{aligned}
        \delta_t &= R_{t+1} \\ &+ \gamma Q^\pi(s_{t+1},a_{t+1};\theta^Q) - Q^\pi(s_t,a_t;\,\theta^Q).
    \end{aligned}
    \end{align}
    Then, the critic update is obtained by:
    \begin{equation}
    \theta^Q_{t+1} \leftarrow \theta^Q_t + \beta \delta_t \nabla_{\theta^Q} Q^\pi(s_t,a_t;\,\theta^Q),
    \end{equation}
    \noindent  where $\beta$ is the learning rate for the critic, and $\gamma$ is the discount factor.
\end{enumerate}
\noindent
The architecture and attributes of the actor applied in the framework of this paper are shown in Fig. \ref{im:NetworksAct}. The actor operates by taking the current state of the environment as input, which includes both the current measurement and past measurements. It then generates a continuous value representing the predicted time to the next merger event, consistent with the continuous action space defined in the environment. The design of the actor model is chosen to capture temporal dependencies within the time series data. This is facilitated by utilizing a long short-term memory (LSTM) layer, which shall effectively learn these dependencies. Layer normalization is applied to normalize inputs across the model to enhance stability during training. The dense (fully connected) layers support the actor in learning the patterns within the data. These layers, equipped with nonlinear activation functions, enable the network to establish connections between the input state and the resulting action output.

The architecture of the critic used in this paper is presented in Fig. \ref{im:NetworksCrit}. The critic serves as the evaluative component in this setup, responsible for assessing the effectiveness of actions taken by the actor model given the current environmental state. It takes two primary inputs: the current state of the environment and the action executed by the actor. The output from the critic model is a scalar value, indicating the expected future cumulative rewards from the current state-action pair, thus offering a direct critique of the actor's decision-making process.
A feature of the critic model is its sequential processing capability, akin to the actor model, which is used for understanding the temporal dynamics leading to the current state. This capability is achieved by incorporating an LSTM layer, enabling the model to evaluate the potential future rewards associated with state-action pairs over time. Additionally, the critic model utilizes dense layers to support learning a complex mapping from state-action pairs to their respective value estimates. These estimates guide the policy improvement of the actor. 

The critic network employs a linear activation function to accommodate positive and negative outputs. Since the actor's action is consistently positive (as to be predicted merger times are always set in the future), we selected ReLU activations.
\newline
\paragraph{Q-Target Networks}
\,\\ \,\\
To enhance the stability of learning, our framework employs Q-target networks for both the actor and critic, denoted as $\theta^{\pi'}$ and $\theta^{Q'}$, respectively. These networks are clones of the actor and critic but with their weights updated less frequently. This approach mitigates the feedback loop issue by providing stable targets for the updates \cite{Zhang2023, Sun2022}:
\begin{equation}
\theta^{Q'}_{t+1} \leftarrow \tau\theta^Q_t + (1 - \tau)\theta^{Q'}_t,
\end{equation}
\begin{equation}
\theta^{\pi'}_{t+1} \leftarrow \tau\theta^\pi_t + (1 - \tau)\theta^{\pi'}_t,
\end{equation}
where $\tau \in (0,1)$ is the soft update coefficient obtained via multi-agent-based hyperparameter optimization.
\newline\newline\noindent 
By weaving together these two principles, the \texttt{Agent} class is engineered to refine its predictive strategy through a balanced process of exploration, leveraging immediate and future rewards. The integration of the actor-critic architecture with target networks facilitates a more robust learning environment.
\newline\newline\noindent 
Finally, the main function call is presented:

\begin{algorithm}
\caption*{Main Function: \texttt{Environment}-\texttt{Agent} interaction}\label{alg:main_function_interaction}
\begin{algorithmic}[1]
\State \texttt{env = Environment, env.initialize(...)}
\State \texttt{agent = Agent, agent.initialize(...)} 
\For{\texttt{episode in range(total episodes)}}
    \State \texttt{state = env.reset()}   
    \For{\texttt{step in range(max. steps per episode)}}
        \State \texttt{action = agent.act(state)} 
        \State \texttt{next state, reward = env.step(action)} 
        \State \texttt{agent.train(state, action, reward, next \text{\,\,\,\,\,\,\,\,\,\,\,\,\,\,\,\,\,\,}state)} 
        \If{\texttt{merger reached}}
            \State \texttt{break} 
        \EndIf
        \State \texttt{state = next state} 
    \EndFor
    \State \texttt{agent.update\_target\_networks()} 
\EndFor
\end{algorithmic}
\end{algorithm}
\noindent
The \texttt{Environment}-\texttt{Agent} interaction is primarily characterized by the agent's \texttt{action}, processed and evaluated by the environment, coupled with the \texttt{state} and action-dependent \texttt{reward} that the environment subsequently provides to the agent.
\subsection{Multi-Agent Reinforcement Learning}\label{subsection:MARL}
Multi-agent reinforcement learning is a subfield of reinforcement learning where multiple agents interact with their environment simultaneously and may interact with each other. Unlike traditional reinforcement learning scenarios with only one agent, multi-agent reinforcement learning involves multiple agents, each making decisions independently while possibly influencing each other's experiences and learning process. In multi-agent reinforcement learning, agents typically share the same environment and may have access to shared or partial information about the experiences of other agents. The objective of each agent may vary, ranging from cooperation to competition or a combination of both \cite{8970977, Albrecht2024-ul}.

\begin{figure}[h!]
	\centering
  \includegraphics[trim=88 130 410 89, clip,width=0.49\textwidth]{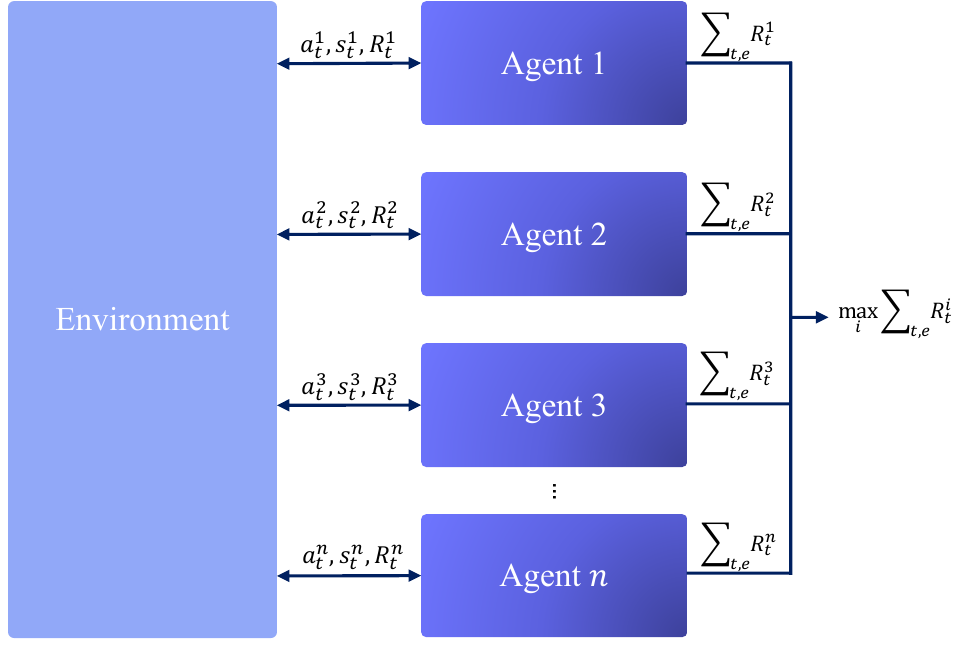}
	\caption{Multi-agent approach used for hyperparameter optimization in predicting massive black hole merger times detected by LISA. The figure showcases the parallel training of multiple agents with distinct hyperparameter configurations across multiple episodes $e$. All agents share the same environment. The optimal agent configuration, determined by the maximum total reward, informs the selection of hyperparameters for the predictive model of the proposed pipeline.}
	\label{im:MARLHyp}
\end{figure}

In this study, a multi-agent hyperparameter optimization approach is implemented to determine the optimal parameters for the reinforcement learning model. This method involves running multiple independent agents in parallel, each with a unique set of hyperparameters randomly drawn from a defined search space. The space includes a range of parameters such as learning rates, discount factors, and parameters for balancing exploration and exploitation. Table \ref{tab:hyperparameters} provides the complete list of parameters. The multi-agent approach is illustrated in Fig. \ref{im:MARLHyp}.

\noindent 
For the optimization, the \texttt{Environment} and \texttt{Agent} classes are implemented as described previously. Throughout the training episodes, agents strive to accurately predict the timing of future black hole mergers using the provided data. The environment's feedback, delivered through rewards, aids in refining their decision-making strategies. The overall performance of each agent is assessed by the total reward accumulated over all episodes.

After completing the trials, the agent configuration that achieves the highest total reward is identified, and its hyperparameters are deemed optimal for the task at hand. This multi-agent hyperparameter optimization leverages parallel processing to expedite the tuning process. The optimization results are provided in Table \ref{tab:hyperparameters}.

\begin{table}[h!]
\captionsetup{justification=centering}
\caption{Overview of hyperparameters obtained from the multi-agent optimization.}
\begin{tabular}{>{\raggedright}p{5.35cm}>{\raggedright\arraybackslash}p{2.65cm}} 
\hline\hline
\textbf{Hyperparameter} & \textbf{Optimization Result}  \\
\hline
Discount factor $\gamma$  & $0.93$ \\
Epsilon $\epsilon(t=0)$, decay rate $\dot\epsilon$  & $1,\,0.9998\,1/\text{steps}$ \\
History size $h$  & $83$ hours \\
Learning rates $\alpha,\,\beta$  & $5\cdot 10^{-7}$ \\
Maximum reward $R_{\text{max}}$ & $90$ \\
Reward sensitivity $w$  & $0.005$ \\
Target network update coefficient $\tau$  & $0.005$ \\
\hline\hline
\end{tabular}
\label{tab:hyperparameters}
\end{table}

\section{\label{sec:level3} Reward Generation with Truncated Waveforms}
The generation of accurate rewards in the absence of real-time merger information poses a significant challenge during testing and evaluation of the reinforcement learning framework. This section focuses on the approach of utilizing truncated waveform models for reward calculation according to Eq. \eqref{eq:reward}. Through template matching with these models, we aim to enable the reinforcement learning agent to adapt its actions dynamically to the observational data when necessary.

\subsection{Template Matching }\label{eq:subsecGWA}

Template matching is a statistical method used to identify the presence of predicted gravitational wave signals in noisy data \cite{PhysRevD.63.044023}. This technique plays an important role in LISA data analysis, where the interferometric data is processed via TDI. 
The use of TDI channels for template matching involves correlating the processed data with a set of precomputed waveforms, so-called templates, to identify potential gravitational wave events. The quality of a match is quantified using the match filter output $\rho$, defined as:
\begin{equation}
\rho = \max_{\theta} \frac{\left\langle d(t) \mid h\left(t,\theta\right)\right\rangle}{\sqrt{\left\langle h\left(t,\theta\right) \mid h\left(t,\theta\right)\right\rangle}},\label{eq:TMPMatch}
\end{equation}
where $d(t)$ represents a TDI channel of interest, $h(t,\theta)$ is a template waveform defined by a set of parameters $\theta$, and $\langle\cdot|\cdot\rangle$ denotes the inner product weighted by the detector's noise spectral density. The goal is to maximize $\rho$, as the template achieving this maximization is considered the best match, providing estimates for the source parameters and the time of merger used for reward generation.

Given the predictive nature of our reinforcement learning framework, we employ truncated IMRPhenomD waveforms that describe the gravitational wave signal during the inspiral, excluding the merger phase and the ringdown. This is due to the fact that the full signal is unavailable during the prediction phase. The truncation process involves cutting off the waveform at $t_{\text{cut}}$, which corresponds to the latest time step available:
\begin{equation}
h_{\text{trunc}}(t) = 
\begin{cases} 
h(t) & \text{if } t \leq t_{\text{cut}}, \\
0 & \text{otherwise}.
\end{cases}
\label{eq:hCut}\end{equation}
Equation \eqref{eq:hCut} is applied to perform template matching according to Eq. \eqref{eq:TMPMatch}. We solve Eq. \eqref{eq:TMPMatch} via differential evolution optimization similar to \cite{StrubGPU, Storn1997}  and use the estimated merger time to generate rewards. 
\newline\newline\noindent
In line with this approach, the \texttt{Environment} class is adapted:
\FloatBarrier
\begin{algorithm}
\caption*{Class Adaption: \texttt{Environment}}\label{alg:Environment_class}
\begin{algorithmic}[1]
\Function{Step}{action}
    \State {Call \textsc{CalculateNoisyReward}.}
    \State {Call \textsc{GetState} to update internal state.}
    \State  \hspace{10pt} \textbf{return} new state and noisy reward
\EndFunction
\Function{CalculateNoisyReward}{action, step}
    \State {Call \Call{TemplateMatching.}{} }
    \State {Calculate noisy reward with template matching result.}
    \State \hspace{10pt} \textbf{return} noisy reward
\EndFunction
\Function{TemplateMatching}{data}
    \State {Correlate right-censored TDI data with truncated \,\,\hspace{15pt}\,\text{\,\,\,\,\,\,\,} IMRPhenomD waveforms.}
    \State {Identify the best matching waveform.}
    \State {Obtain estimated merger time from the best match.}
    \State \hspace{10pt} \textbf{return} estimated merger time
\EndFunction
\end{algorithmic}
\end{algorithm}
\subsection{Handling Noisy Rewards in the Reinforcement Learning Framework}\label{subsec:NoisyRew}
The introduction of noise in the reward signal, especially due to uncertainties in waveform template matching, poses a considerable challenge in our predictive framework. This complication arises because the reward signal serves as the critical feedback mechanism that guides the agent to adapt to the data when adaption is necessary. When the reward signal is perturbed or noisy, it can significantly disrupt the agent's ability to learn optimal behaviors. Drawing upon \cite{Wang2020, EverittRL2017, majadas2021disturbing}, we analyze the impact of reward perturbation and outline strategies to mitigate its effects on the quality of the agent's actions.

\subsubsection{Impact of Noisy Rewards}
Noisy rewards, or reward perturbation, pose a significant challenge to the learning efficiency and policy performance of reinforcement learning agents. Even minor perturbations in the reward signal can induce substantial deviations in the policies learned by the agents. These deviations can lead to suboptimal or erroneous action selections, undermining the agent's performance. In the domain of this paper, such inaccuracies are particularly problematic as they directly translate into errors of time-to-merger predictions. The noise in the reward signal jeopardizes the convergence stability and the optimality of the learned policy, underscoring the imperative for deploying effective noise management strategies within the reinforcement learning framework.

Ideally, the reward signal should accurately reflect the quality of actions towards achieving the agent's objectives. However, in practical scenarios, the reward signal often gets contaminated with noise, leading to a noisy reward, $R_{t,\text{noisy}}$. This can be formalized as:
\begin{equation}
    R_{t,\text{noisy}} = R_{t,\text{ideal}} + \Delta R_t,
\end{equation}
where $\Delta R_t$ represents the noise component due to uncertainties in the reward generation at time step $t$.

\noindent
This addition of noise to the reward signal leads to discrepancies in the action selection process, affecting the agent's performance. The deviation in action selection caused by the noisy reward can be represented as:
\begin{equation}
    \delta a_t = a_{t,\text{noisy}} - a_{t,\text{ideal}},\label{eq:deltaAct}
\end{equation}
where $a_{t,\text{noisy}}$ is the action chosen based on the noisy reward, and $a_{t,\text{ideal}}$ is the action that would have been selected under ideal, noise-free conditions at time step $t$.

Minimizing the deviation $\delta a_t$ is crucial for maintaining the accuracy of the agent's decision-making process. This involves developing and implementing effective mitigation strategies to reduce the impact of $\Delta R_t$, thereby ensuring that the agent's actions remain closely aligned with its objectives, even in the face of reward uncertainty.

\subsubsection{Mitigation Strategies}
\begin{figure}[]
	\centering
  \includegraphics[trim=85 280 390 40, clip,width=0.5\textwidth]{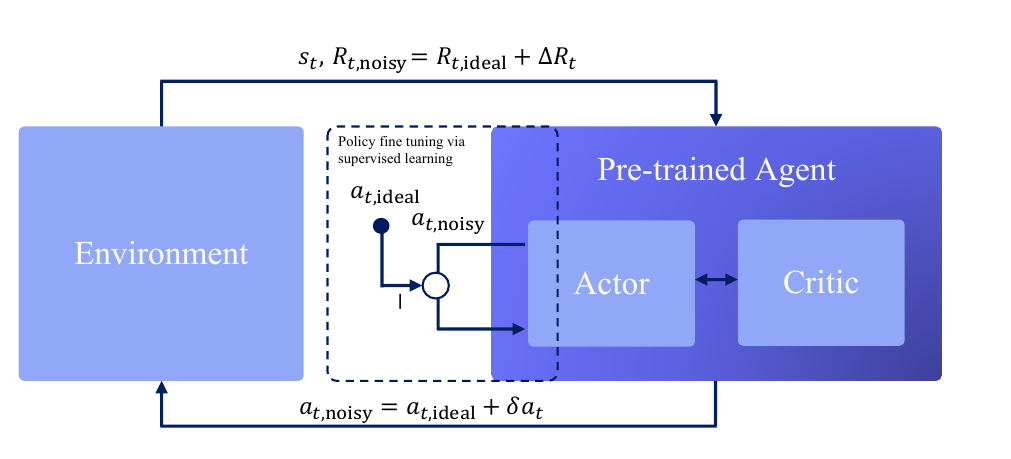}
	\caption{Policy fine-tuning in the reinforcement learning model. Stage 2 of training illustrates the integration of supervised learning to update the actor's weights, aiming to preserve ideal action choices in the presence of perturbed rewards.}
	\label{im:SLPFT}
\end{figure}

To counteract the adverse effects of perturbed rewards,  different mitigation strategies exist that directly address the noise $\Delta R_t$ in the reward signal, its effects $\delta a_t$ on the agent's actions, or focus on enhancing the inherent robustness of the reinforcement learning network by dampening the influence of reward perturbations on the prediction process. Three strategies are outlined in the following: 
\newline\newline\noindent
1. Reward filtering: Implementing a reward filtering mechanism aims to smooth out the noise in the reward signal through statistical techniques, such as moving averages. By applying a filter to the reward signal, the reinforcement learning agent receives a more stable and consistent training signal that can enhance learning efficiency and policy reliability. Additionally, incorporating error estimation mechanisms through auxiliary networks can further refine the reward signal by predicting and compensating for the anticipated noise, thereby reducing $\Delta R_t$. This approach requires knowledge about the expected noise level in the reward signal and its evolution over time.
\newline\newline\noindent
2. Adaptive learning rate adjustment:
Adapting the learning rate based on the perceived noise level in the reward signal is a dynamic approach to mitigate the impact of perturbations. By lowering the learning rate in response to high noise levels, the reinforcement learning model can become more conservative in updating its policy, thereby preventing rapid divergences from the optimal policy. Conversely, when noise levels are low, the learning rate can be increased to accelerate policy refinement. The effectiveness of this approach also benefits from knowledge about the noise characteristics.
\newline\newline\noindent
3. Policy-fine tuning:
A measure explored in this paper is the implementation of supervised learning-based policy fine-tuning during training. This approach leverages a dataset of ideal actions to guide the reinforcement learning agent toward more accurate decision-making despite noise in the environment's feedback. By adjusting the policy in light of known ideal actions, the agent is supposed to reduce $\delta a_t$ in the presence of compromised rewards.
\newline\newline\noindent
To refine the predictive capabilities of our reinforcement learning model, we employ a two-stage training process that incorporates the policy fine-tuning strategy. In the initial stage, the focus is on leveraging $R_{t,\text{ideal}}$ to achieve a pre-trained agent capable of making reliable predictions, considering noisy TDI data against a backdrop of ideal rewards. This foundational training phase is important for establishing a baseline of accuracy and reliability in the agent's predictive behavior. Subsequently, we transition to the second training stage, dedicated to policy fine-tuning, which supports enhancing the model's resilience to reward perturbations $\Delta R_t$. This stage is characterized by a blend of reinforcement learning and supervised learning. Specifically, the reinforcement learning process alternates with supervised learning phases that adjust the weights of the actor model via back-propagation. During supervised learning, the agent is presented with perturbed rewards, and the actor's model parameters are updated such that $\delta a_t$ of Eq. \eqref{eq:deltaAct} is minimized. The principle is illustrated in Fig. \ref{im:SLPFT}.
\,
\newline\newline\noindent
The modified \texttt{train} function of the \texttt{Agent} class is shown on the next page. It is used for both training stages. Supervised policy fine-tuning is only applied if \texttt{train} is provided with the ideal action $a_{t,\text{ideal}}$.
\FloatBarrier
\begin{algorithm}
\caption*{Function: \texttt{train} within \texttt{Agent}}
\begin{algorithmic}[1]
\Function{Train}{state, action, noisy reward, next state, ideal action}
    \State {Predict next action using target actor model.}
        \If{ideal action is provided}
        \State {Calculate loss between predicted and ideal actions.}
        \State {Compute gradients for the actor model.}
        \State {Update actor model using supervised learning gra-\text{\,\,\,\,\,\,\,\,\,\,\,\,\,\,\,\,\,\,}dients.}
    \EndIf
    \State {Calculate future Q-value from target critic model.}
    \State {Compute target Q-value incorporating discount factor \text{\,\,\,\,\,\,\,\,\,}and reward.}
    \State {Obtain current Q-value from critic model.}
    \State {Calculate critic and actor losses.}
    \State {Compute gradients for both actor and critic models.}
    \State {Apply gradient clipping to stabilize training.}
    \State {Update actor and critic models with computed \text{\,\,\,\,\,\,\,\,\,}gradients.}
\EndFunction
\end{algorithmic}
\end{algorithm}
\vspace{-10pt}
\subsection{Simulation Results}
The subsection demonstrates the performance of the reinforcement learning framework in predicting the future merger times of massive black hole binaries in LISA data. Initially, the simulation setup is presented.
\vspace{0pt}
\subsubsection{Simulation Setup}
\vspace{-8pt}
The simulation environment is designed to meet the expected observational conditions of LISA. The dataset utilized is composed of synthetic waveforms that are generated and superimposed using the IMRPhenomD model. Signal strengths and source parameters are randomly sampled from the astrophysical catalog employed by the LISA Spritz dataset \cite{SpritzLDF}. We consider distances spanning up to 100 Gpc and component masses ranging from $10^6$ to $10^7\,\mathrm{M}_{\odot}$. Additionally, instrumental noise is considered in accordance with the specifications outlined in the LISA science requirement document \cite{SciReqDoc}. The cadence set for the reinforcement learning framework is established at 60 seconds, while for the truncated template matching, it is set at 5 seconds, aligning with the cadence of the LISA Spritz dataset. Analog to \cite{SpritzLDF}, Keplerian LISA orbits are considered optimized to minimize variations in light travel time up to the second order in the orbital parameter.

We incorporate variable intervals between consecutive mergers, randomly ranging from 20 to 100 hours. Estimates indicate that LISA has the potential to detect 1 to 20 massive black hole mergers annually \cite{strub2024global, PhysRevLett.117.101102, 10.1093/mnras/stz3102}. The intervals between merger events can thus be significantly larger in reality. However, this choice is made to control the total volume of simulated data. In one sense, the reinforcement learning agent encounters a more challenging scenario as a result, as additional mergers lead to increased source confusion through their superposition in TDI.
To ensure the agent's adaptability to real measurements, characterized by larger periods between consecutive mergers, we continuously execute the convolutional neural network of Section \ref{sec:levelDet} while the agent remains in standby mode. The agent with a maximum action space of 100 hours is activated when a signal has been detected and the variance in the time-to-merger prediction obtained with truncated template matching decreases consistently for the first time. Then, the reinforcement learning framework proceeds through episodes, each comprising a varying amount of time steps where the agent interacts with the environment to improve its policy. A new episode begins upon detecting a merger, with the past observation buffer refreshed accordingly. 

The implementation differentiates between the training and testing phases. Training occurs in two distinct steps. Initially, ideal rewards are employed to pre-train the agent using hyperparameters derived from multi-agent optimization, utilizing 25,000 merger events. The training dataset is constructed without relying on prior knowledge of astrophysical parameter distribution, opting for a balanced representation across the parameter space to mitigate prediction biases and enhance generalizability. We assessed the impact of varying training sizes on testing performance, focusing on identifying a threshold beyond which no significant improvements were observed. It shall be noted that, in general, reinforcement learning does not rely on a predefined training set. Instead, a reinforcement learning agent learns continuously from interactions with an environment, which provides feedback through rewards or penalties. Since agents learn from experience, the amount of data they can learn from is potentially infinite. The agent can keep interacting with the environment, gathering more experience and refining its policy accordingly. The training set size mentioned here is to be understood as the number of merger events, after which the second dedicated training phase is started. In the second step, the policy of the pre-trained agent is further refined using noisy rewards and the fine-tuning approach outlined in Section \ref{subsec:NoisyRew}, utilizing another 25,000 merger events. Our study shows optimal performance when applying the supervised learning approach only within the initial hours of each prediction episode. Moreover, policy fine-tuning is reapplied periodically every 50 merger events during testing phase. These episodes are excluded from the testing data evaluation, and 5,000 test events are assessed in total.

For this initial framework testing, it is presupposed that LISA benefits from uninterrupted data updates, a simplification that facilitates data handling. In reality, communication within the constellation will be performed via a designated spacecraft, a relay node, over 5 days. During this interval, connectivity is maintained for 8 hours daily. This operational aspect is neglected in the beginning.
\vspace{-10pt}
\subsubsection{Prediction Performance}\label{subsub:Pred}
\begin{figure}[]
    \centering
        \centering
        \includegraphics[trim=30 0 0 37, clip, width=0.52\textwidth]{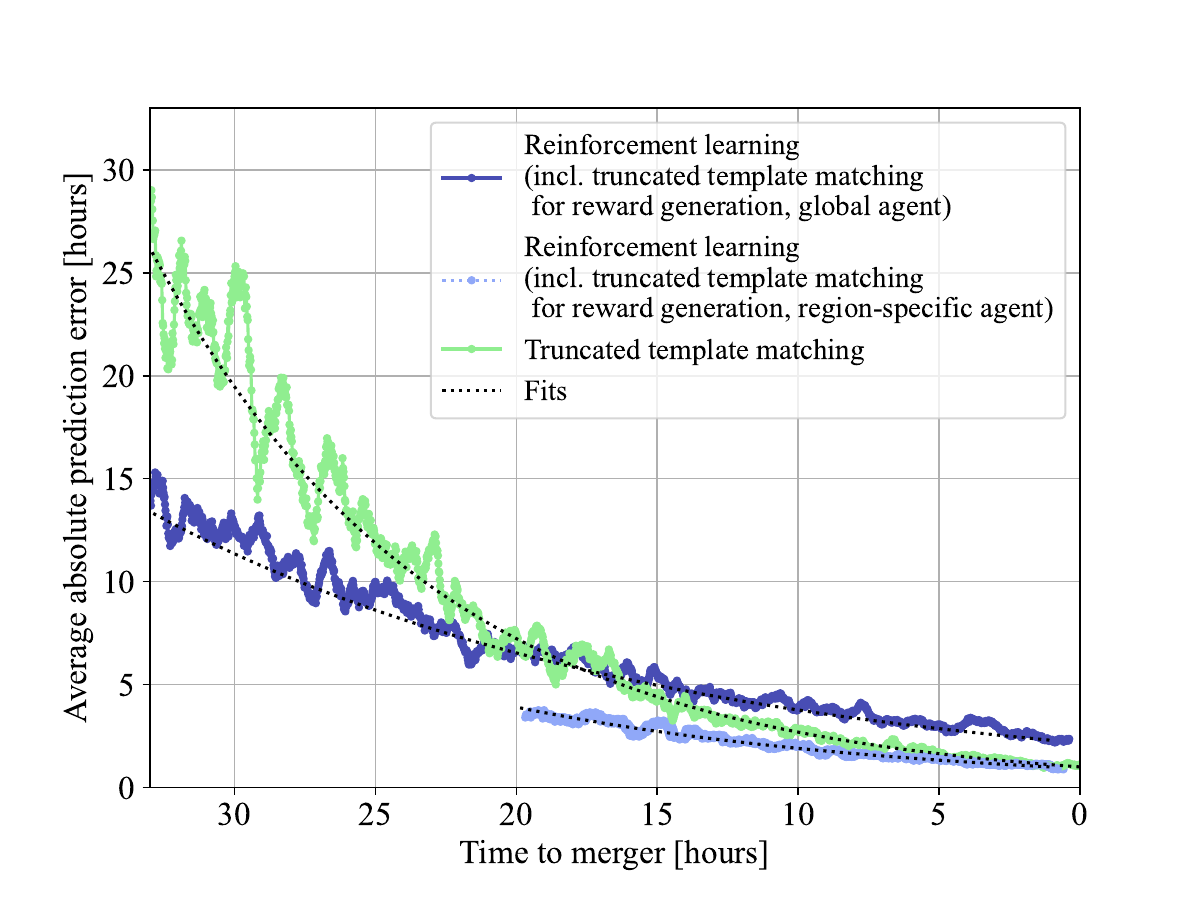}
        \captionsetup{}

    \caption{Prediction performance. The plot shows the average absolute merger time prediction error for the described testing set, once for the case where truncated template matching is executed alone and once for the reinforcement learning framework, which utilizes the result of truncated template matching for reward generation. In the context of the reinforcement learning approach, two distinct agents are employed. The 'global agent' undergoes training covering the entire action space. Conversely, the 'region-specific agent' receives specialized training tailored to a narrower range, specifically up to 20 hours pre-merger, and a lower learning rate. We concentrate on the time interval wherein the absolute prediction error of the global agents diminishes to less than half of the actual time to merger. Beyond this interval, the outputs of the reinforcement learning agent would still remain valuable for detecting a signal, which marks the initial step in the proposed prototype pipeline. In this paper, we employ a convolutional neural network for the detection task. This is due to its ability to operate without the need to specify additional detection thresholds.}
    
    \label{fig:AvPredictionError}
\end{figure}

\vspace{-8pt}
The time-to-merger prediction performance is depicted in Fig. \ref{fig:AvPredictionError}. The plot illustrates the average absolute prediction error derived from the testing set under two scenarios: employing truncated template matching in isolation and integrating it into the reinforcement learning framework, as described, where truncated template matching is utilized for reward generation. In the context of the reinforcement learning approach, two distinct agents are employed. The global agent undergoes training covering the entire action space up to a maximum of 100 hours of merger events stretching into the future. Conversely, the region-specific agent receives specialized training tailored to a narrower range, specifically up to 20 hours pre-merger. The associated action space is reduced accordingly. The term 'average' denotes averaging over all massive black hole merger events without classifying them based on factors such as redshift or component masses. This approach provides a holistic view of the overall performance across the entire parameter space. A more distinct analysis will be conducted in the next section, focusing on an individual signal with a defined signal-to-noise level. We concentrate on the time interval wherein the absolute prediction error of the global agents diminishes to less than half of the actual time to merger.

Figure \ref{fig:AvPredictionError} reveals the superiority of the global agent up to approximately 16 hours before the merger event compared to truncated template matching. Below this temporal threshold, executing truncated template matching alone minimizes the prediction error in direct comparison. While a comparable performance with the global agent could be achieved within this period, it necessitates selecting higher learning rates for the actor and critic networks. With increasing learning rates, the agent's prediction performance gradually converged to the template-matching performance. This can be attributed to the increasing emphasis placed on rewards derived from truncated template matching within the reinforcement learning framework. The convergence then implies a decline of the agent's prediction performance for time intervals surpassing 16 hours before the merger event.  Hence, the challenge in devising an accurate agent lies in selecting suitable learning rates to obtain the best prediction from internal knowledge of the network based on historical experience and adapting to the new situation according to the rewards received.

An advancement of the current framework could involve implementing an adaptive learning rate mechanism, which dynamically adjusts based on the agent's current prediction. In this method, the learning rate would initially fall within the range specified in Table \ref{tab:hyperparameters} for time periods exceeding the 16-hour threshold, gradually increasing thereafter. Suitable changes to the learning rate could also be determined through multi-agent hyperparameter optimization as performed in Section \ref{subsection:MARL}.

\begin{figure*}[]
        \centering
        \includegraphics[trim=0 0 130 160, clip, width=0.69\textwidth]{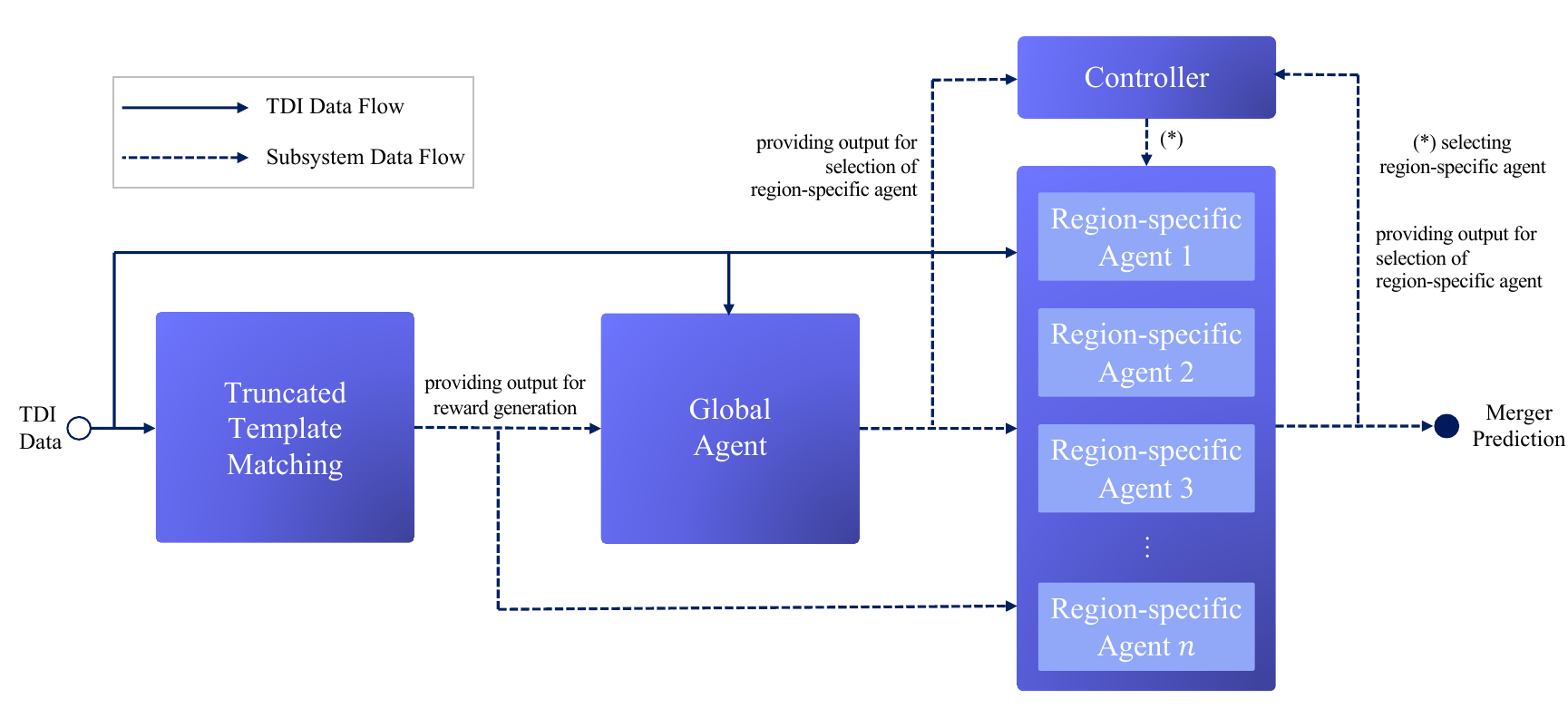}
    \caption{The diagram illustrates the sequential refinement of merger predictions involving global and region-specific reinforcement learning agents. TDI data enters from the left and is processed first by truncated template matching. The output is then utilized for reward generation and directed into two distinct pathways, one leading to the global agent and the other to a series of region-specific agents. The global agent is trained with a large action space, while the region-specific agents are trained with narrower action spaces tailored to specific pre-merger temporal intervals, such as 0 hours to 20 hours and 20 hours to 40 hours per-merger, along with the associated data. The predictions from both the global agent and the active region-specific agent are channeled to a controller entity. The controller determines which region-specific agent is most appropriate to activate at any given time, knowing the action space characteristics of the various region-specific agents. This sequential activation of agents, guided by the controller's decisions, aims to enhance the accuracy of the merger time prediction, concluding with the final and refined merger prediction. Future investigations will explore the optimal number $n$ of region-specific agents as well as the impact of increasing agent density with decreasing action spaces as the merger event approaches.}
    \label{fig:Flowchart_Agents}
\end{figure*}

Another approach involves the implementation of multiple region-specific agents. For instance, one agent is trained on data up to 20 hours pre-merger, while a second agent is trained on the region spanning from 20 hours to 40 hours, and so forth. These agents can then be executed sequentially, with transitions from one agent to the next determined by the predictions of the last active local agent. The concept is outlined in Fig. \ref{fig:Flowchart_Agents}.
Further investigation into the design of these transitions is required to ensure their smoothness, although this falls outside the scope of this paper. The paper demonstrates the performance of a region-specific agent operating within a 20-hour time frame instead of 100 hours of the global agent, as depicted in Fig. \ref{fig:AvPredictionError}. The results exhibit an improvement within this time frame compared to the global agent, with reduced variance. Note that the merger and a few minutes leading up to it are omitted from the training and testing. This exclusion is due to the significant differences between the close-to-merger and the inspiral phases and supports the prediction performance of the agents.

Note that the reinforcement learning setup could also trigger alerts, aligning with the first task in the prototype pipeline. Despite the prediction performance decreasing as the time-to-merger increases, the agent's predictions retain their value for the detection task. Nevertheless, we selected the convolutional neural network for detecting massive black hole binary signals primarily because of its capacity to operate without the necessity of specifying additional detection thresholds.\enlargethispage{5\baselineskip}
\vspace{0pt} 
\section{Validation with the LISA Spritz Dataset}\label{sec:level5} 
\vspace{-5pt}
The LISA data challenges aim to replicate the observational conditions of LISA, providing an opportunity to develop, refine, and test data analysis techniques crucial for interpreting gravitational waves before the mission launch. These challenges generate realistic TDI data, covering a wide range of gravitational wave sources such as massive black hole mergers and galactic binaries, while also incorporating expected instrumental noise like shot noise and test mass acceleration noise. The datasets provided enable a direct comparison of different algorithm performances, which is essential for identifying effective data analysis methods. For validation purposes, the framework of this paper is applied to the dataset that contains a prominent massive black hole merger signal with a high signal-to-noise ratio of approximately 2000. The aim is to detect the signal during its early inspiral phase and to predict the merger time before culmination. 

\begin{figure}[]
    \centering
        \centering
        \includegraphics[trim=20 0 0 37, clip, width=0.51\textwidth]{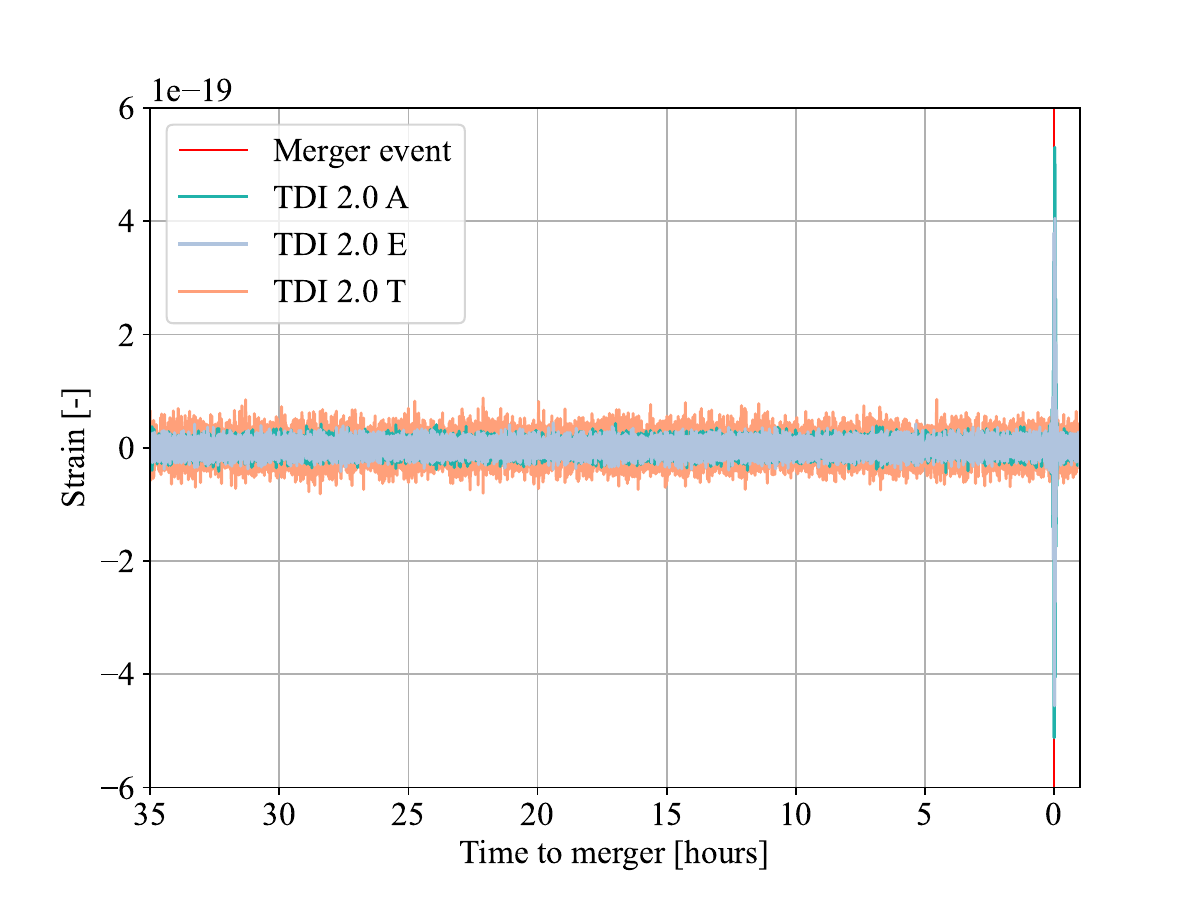}
    \caption{Time series data of the TDI 2.0 variables A, E, and T for the massive black hole binary merger extracted from the LISA Spritz dataset (glitches removed) with $m_{{1}}=1.32\cdot10^{6}\,$M$_{{\odot}}\,$, and $q=0.463$. The curves illustrate the evolution of these variables over time, with the inspiral phase buried in the noise. The spectrogram of the TDI 2.0 variable A is depicted in Fig. \ref{fig:Spec1_A}.}
    \label{fig:Merger1}
\end{figure}

\begin{figure}[]
    \centering
        \centering
        \includegraphics[trim=15 0 0 37, clip, width=0.52\textwidth]{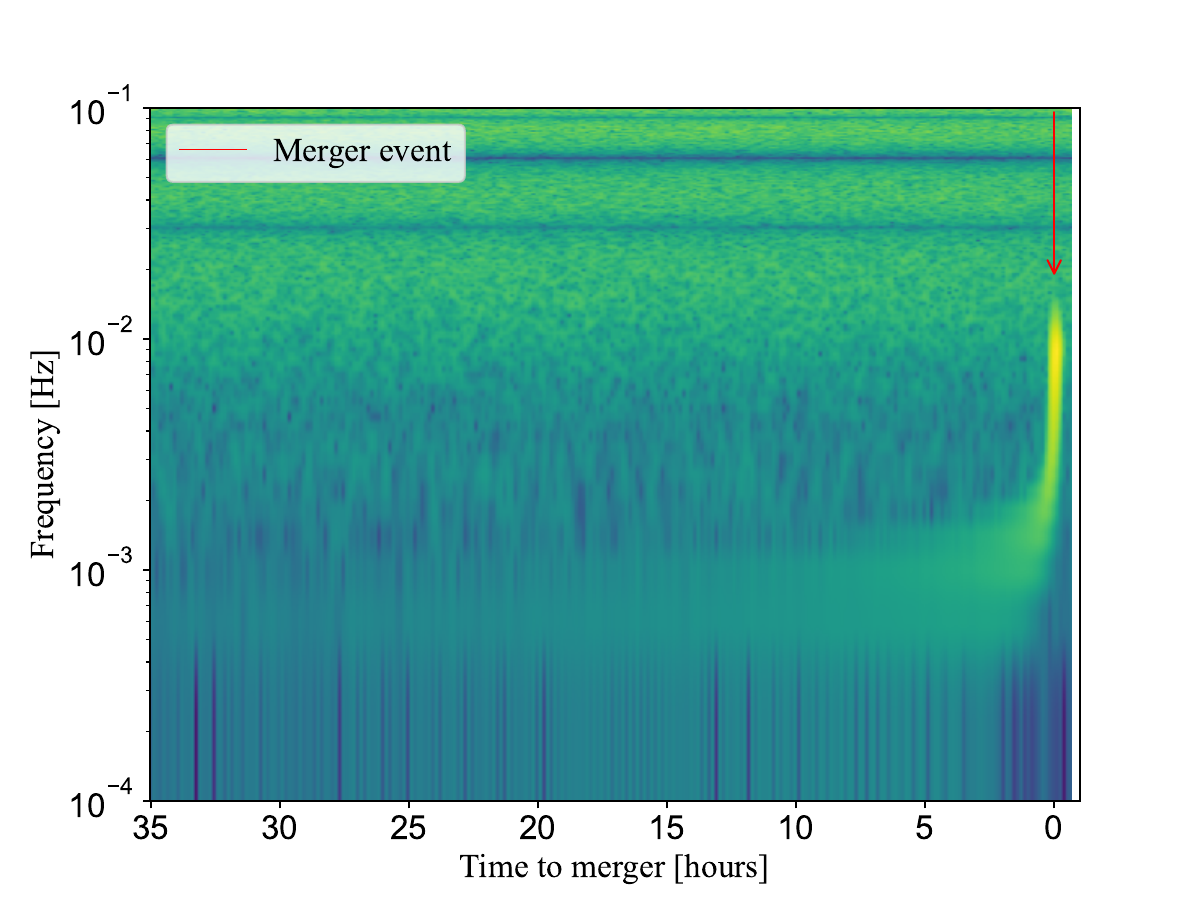}
    \caption{Spectrogram of the TDI 2.0 variable A for the massive black hole binary merger of Fig. \ref{fig:Merger1}. The spectrograms of the TDI 2.0 variables E and T are given by Figs. \ref{fig:Spec1_E} and \ref{fig:Spec1_T} in the appendix.}
    \label{fig:Spec1_A}
\end{figure}

The time series for the TDI second generation (TDI 2.0) variables A, E, and T of the signal under consideration is depicted in Fig. \ref{fig:Merger1}. The corresponding spectrogram of the TDI 2.0 A channel is given in Fig. \ref{fig:Spec1_A}. 

\begin{figure}[]
	\centering
  \includegraphics[trim=5 20 0 0, clip,width=0.48\textwidth]{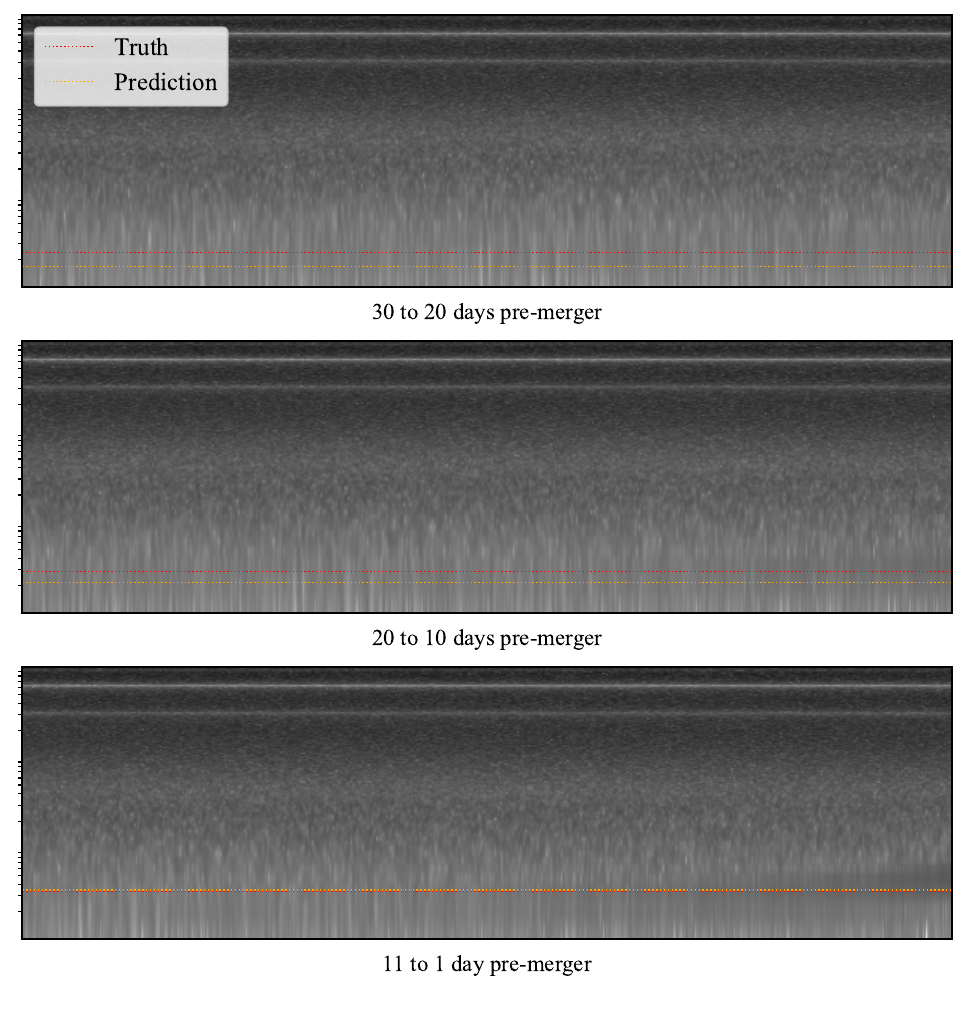}
	\caption{Signal detection in the LISA Spritz dataset as the first step of the prototype pipeline performed with the convolutional neural network presented in Section  \ref{sec:levelDet}. The network identifies the massive black hole binary signal within the LISA Spritz dataset approximately 20 days preceding the merger event, and reliably maintains detection thereafter.} \label{im:TimeEvolvTDISpecSpritz}
\end{figure}

Figure \ref{im:TimeEvolvTDISpecSpritz} illustrates the signal detection result. Signal detection marks the first step in the prototype pipeline. The neural network outlined in Section \ref{sec:levelDet} is applied. The network identifies the merger within the time-evolving TDI spectrograms approximately 20 days before the actual event.

\noindent
Note that glitches and gaps have been eliminated in the data for this initial study. The approach developed here is intended to be integrated with the pipeline from \cite{houba2024detection} in the future to accommodate a more realistic setup.

\begin{figure}[]
    \centering
        \centering
        \includegraphics[trim=20 0 0 0, clip, width=0.465\textwidth]{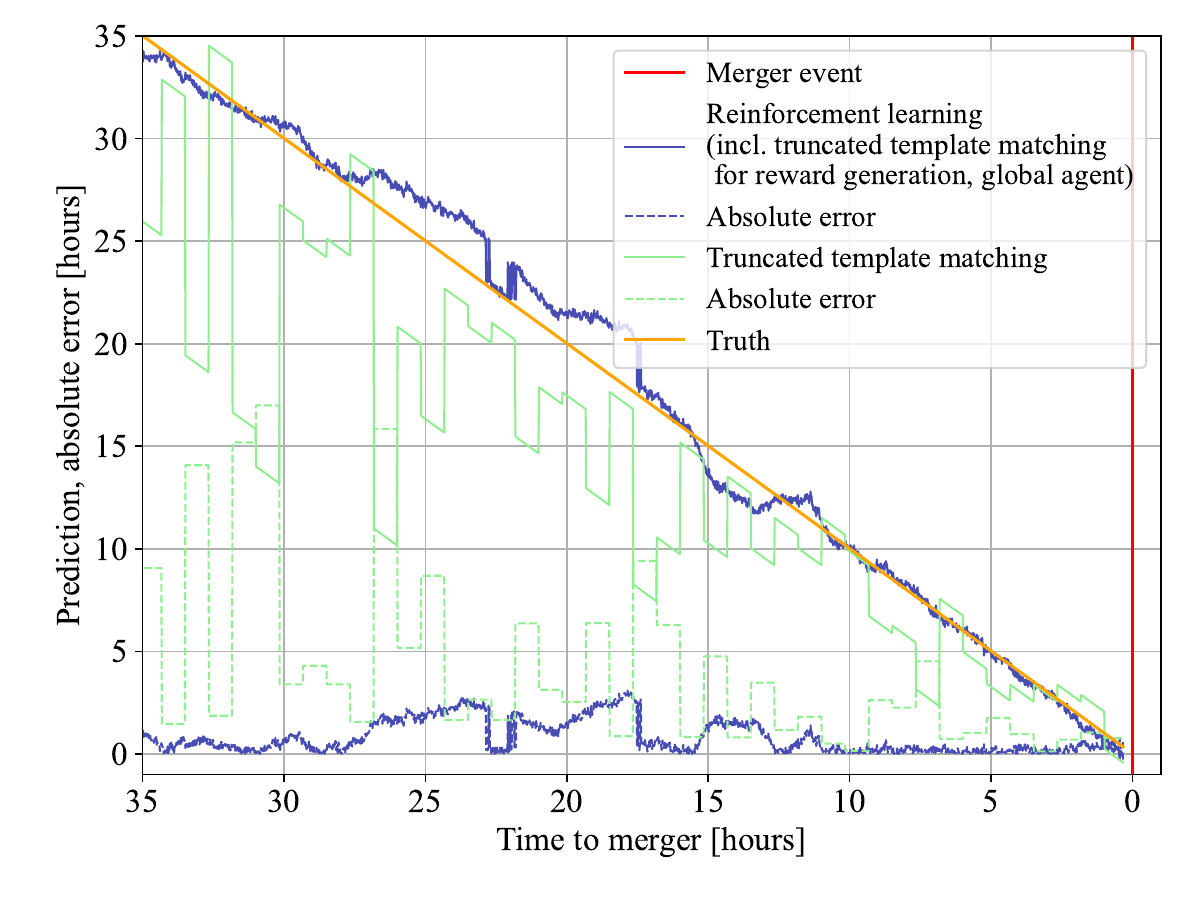}
    \caption{Time-to-merger prediction as the second step of the prototype pipeline, including absolute prediction errors for the example of Fig. \ref{fig:Merger1} applying truncated template matching and the reinforcement learning framework that utilizes the truncated template matching results for reward generation. Here, the reinforcement learning framework predicts the time to merger with reduced absolute prediction errors, thus highlighting its potential for future low latency and alert pipelines in LISA.}
\label{fig:TimeToMerger_Spritz2_33}
\end{figure}

Figure \ref{fig:TimeToMerger_Spritz2_33} displays the associated time-to-merger predictions, including the absolute prediction errors obtained with via reinforcement learning. The figure displays the results again in two instances: once for the case where truncated template matching is applied alone and once for the reinforcement learning framework that utilizes the truncated template matching result for reward generation. The matching algorithm operates only every 50 minutes, resulting in the distinctive step shape observed in the corresponding curves. This approach is used to conserve computational resources. It should be highlighted that the agent's prediction is not merely a filtered version of the matching results; rather, through two distinct training stages, the actor and critic networks effectively approximate the true time to the merger.
 
Figure \ref{fig:TimeToMerger_Spritz2_33} is reprinted in the appendix by Fig. \ref{fig:TimeToMerger_Spritz2_92} for a longer pre-merger period. Figure \ref{fig:TimeToMerger_Spritz2_92} includes 80 hours pre-merger and visualizes the agent's initialization behavior.

\vspace{-10pt}
\section{Conclusion}\label{sec:level6}
We presented a novel prototype pipeline tailored for detecting and predicting future merger times for massive black hole binary systems, particularly in preparation for the Laser Interferometer Space Antenna (LISA) mission. This pipeline operates in two stages. Initially, the pipeline employs a convolutional neural network to detect signals from massive black hole binaries within time-evolving Time-Delay Interferometry (TDI) spectrograms. Subsequently, a reinforcement learning framework is leveraged to predict the time leading up to the merger event. By combining supervised learning, deep Q-learning, and truncated waveform template matching for reward generation, the early warning algorithm demonstrates the potential to be included in future low latency and alert pipelines required for LISA. The actor and critic networks adeptly learn the temporal patterns present in the TDI data by employing a two-stage training process, which incorporates supervised learning-based policy fine-tuning to address the issue of contaminated rewards. This method allows us to estimate the time until the merger across the anticipated range of source parameters. A first validation is performed based on the LISA Spritz dataset.

Beyond the presented framework, incorporating additional transient event types and predicting sky localization remain important objectives for a low latency and alert pipeline. Depending on the event rate, the effect of cross-talk might also emerge as a critical consideration in the future, potentially influencing the precision and reliability of alert information. Integrating data from the global fit will be required for foreground noise subtraction and enhancing prediction performance. The topics will be covered in follow-up research.

It needs to be determined how the detection and prediction performance changes if the signals additionally contain subdominant modes, which are not accounted for in the current waveform model.

Additional areas of investigation arise in the context of reinforcement learning: Deriving the optimal number of region-specific agents remains an open task. The challenge will also be in creating smooth transitions between the predictions of these agents. Moreover, introducing a dynamic learning rate, adjusted based on the context of predictions or observed performance metrics, could optimize the agent's performance, making the system more responsive to data complexities and inherent uncertainties. Another direction for future research involves the introduction of an initial offset in the agent's predictions. This could potentially shorten the initialization phase and enhance the accuracy of predictions in early phases. Additionally, refining noise management strategies in reward processing is expected further improve the fidelity of the reward signal.
\vspace{-5pt}
\begin{acknowledgments}
\vspace{-5pt}
The authors express their gratitude to Cecilio Garcia Quiros and Philippe Jetzer for their valuable feedback on the article. The authors acknowledge support from GW-Learn, a project funded through a Sinergia grant from the Swiss National Science Foundation. The authors thank the LISA data challenge working group for providing the LISA Spritz dataset. The authors acknowledge access to the Euler cluster at ETH Zurich.\par\nopagebreak
\enlargethispage{\baselineskip}
\end{acknowledgments}
\appendix
\section*{Appendix}
\begin{figure}[b!]
    \centering
        \centering
        \includegraphics[trim=10 0 0 37,, clip, width=0.52\textwidth]{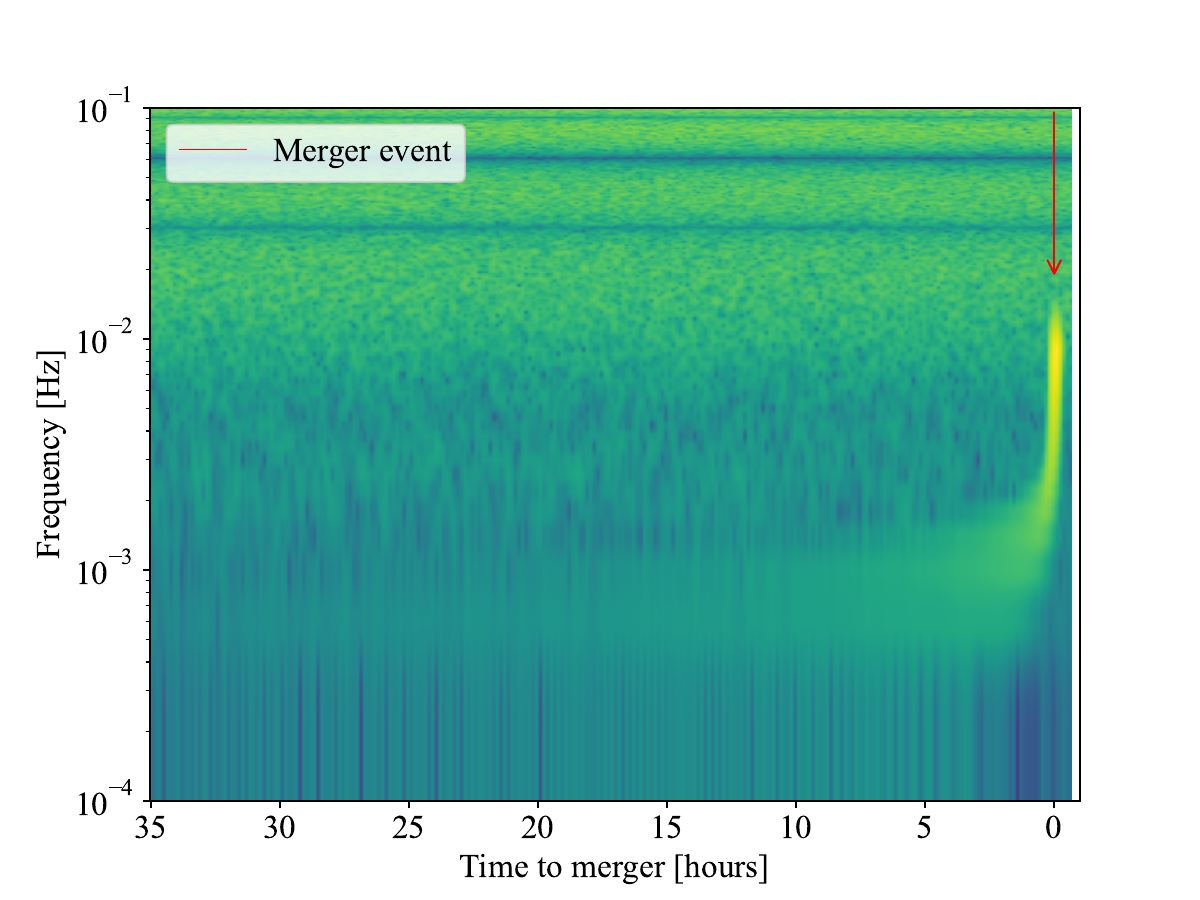}
    \caption{Spectrogram of the TDI 2.0 variable E for the massive black hole binary merger of Fig. \ref{fig:Merger1}.}
    \label{fig:Spec1_E}
\vspace{15pt}
    \centering
        \centering
        \includegraphics[trim=10 0 0 37,, clip, width=0.525\textwidth]{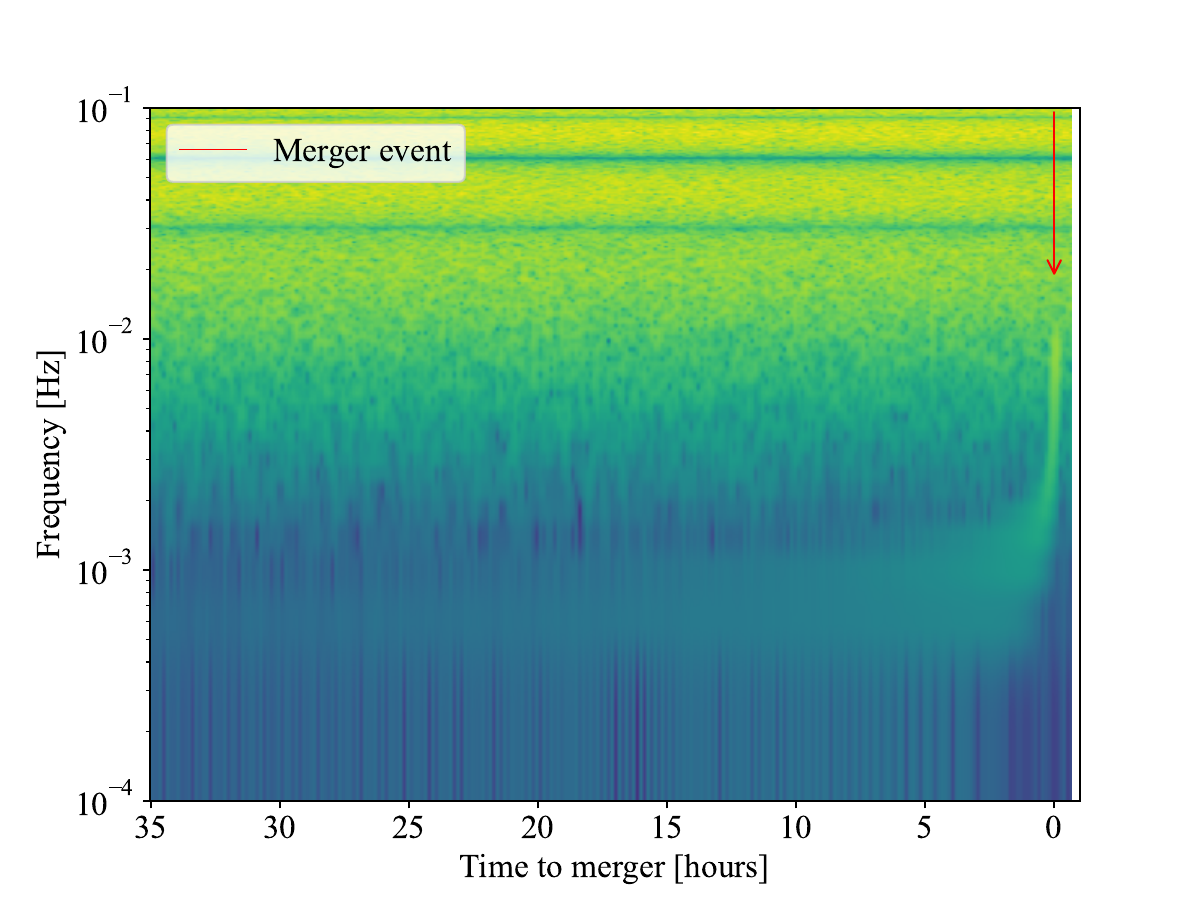}
    \caption{Spectrogram of the TDI 2.0 variable T for the massive black hole binary merger of Fig. \ref{fig:Merger1}.}
    \label{fig:Spec1_T}
\end{figure}

\begin{figure}[]
    \centering
        \centering
        \includegraphics[trim=0 0 0 3, clip, width=0.49\textwidth]{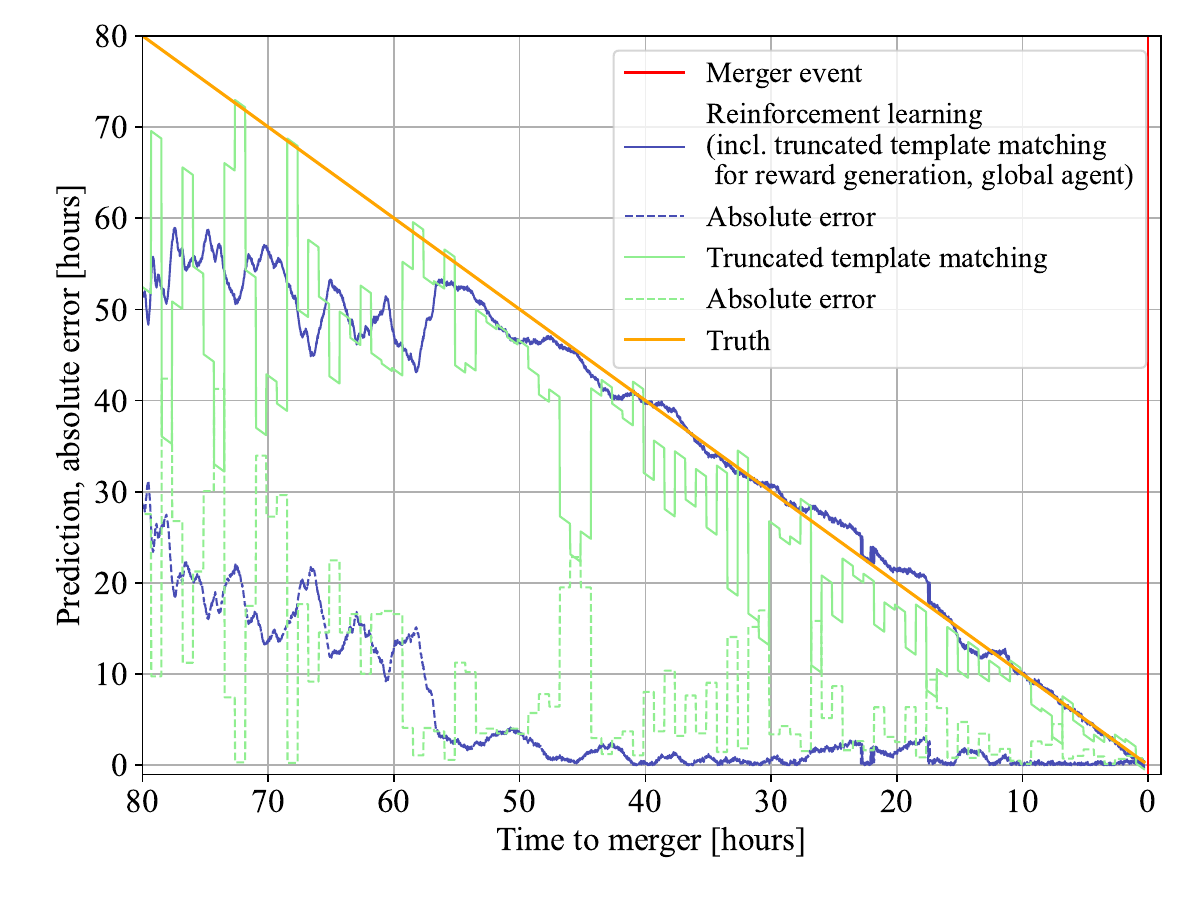}
    \caption{Time-to-merger predictions and absolute prediction errors for the massive black hole binary merger of Fig. \ref{fig:Merger1} applying truncated template matching and the reinforcement learning framework that utilizes the truncated template matching results for reward generation. This figure is equal to Fig. \ref{fig:TimeToMerger_Spritz2_33}, but it spans a longer pre-merger period to visualize the agent's initialization behavior. }
    \label{fig:TimeToMerger_Spritz2_92}
\end{figure}

Figures \ref{fig:Spec1_E} and \ref{fig:Spec1_T} show the spectrograms of the TDI 2.0 variables E and T for the massive black hole binary merger of Fig. \ref{fig:Merger1}. Figure \ref{fig:TimeToMerger_Spritz2_92} represents an extended version of Fig. \ref{fig:TimeToMerger_Spritz2_33} for a longer pre-merger period to visualize the agent's initialization behavior.
\bibliography{apssamp}
\end{document}